# Functional Connectivity of the Brain Across Rodents and Humans

Nan Xu[1], Theodore J. LaGrow[3], Nmachi Anumba[1], Azalea Lee[2,4], Xiaodi Zhang[1], Behnaz Yousefi[1], Yasmine Bassil[2], Gloria Perrin Clavijo[1], Vahid Khalilzad Sharghi[1], Eric Maltbie[1], Lisa Meyer-Baese[1], Maysam Nezafati[1], Wen-Ju Pan[1], Shella Keilholz[1,2]

[1]Biomedical Engineering, Emory University/Georgia Tech
[2]Neuroscience Program, Emory University
[3]Electrical and Computer Engineering, Georgia Tech
[4]Emory University School of Medicine

***Abstract***:

Resting-state functional magnetic resonance imaging (rs-fMRI), which measures the spontaneous fluctuations in the blood oxygen level-dependent (BOLD) signal, is increasingly utilized for the investigation of the brain's physiological and pathological functional activity. Rodents, as a typical animal model in neuroscience, play an important role in the studies that examine the neuronal processes that underpin the spontaneous fluctuations in the BOLD signal and the functional connectivity that results. Translating this knowledge from rodents to humans requires a basic knowledge of the similarities and differences across species in terms of both the BOLD signal fluctuations and the resulting functional connectivity. This review begins by examining similarities and differences in anatomical features, acquisition parameters, and preprocessing techniques, as factors that contribute to functional connectivity. Homologous functional networks are compared across species, and aspects of the BOLD fluctuations such as the topography of the global signal and the relationship between structural and functional connectivity are examined. Time-varying features of functional connectivity, obtained by sliding windowed approaches, quasi-periodic patterns, and coactivation patterns, are compared across species. Applications demonstrating the use of rs-fMRI as a translational tool for cross-species analysis are discussed, with an emphasis on neurological and psychiatric disorders. Finally, open questions are presented to encapsulate the future direction of the field.

## I.     Introduction

Resting-state functional magnetic resonance imaging (rs-fMRI), which detects the spontaneous fluctuations in the blood oxygen level-dependent (BOLD) signal, is a powerful, noninvasive tool for the investigation of the brain's intrinsic functional organization at the macroscale level. Functional connectivity of the brain can be calculated by finding the Pearson correlation between the BOLD signals of different brain areas. This results in spatial maps reflective of the brain's intrinsic functional organization (B. Biswal et al., 1995). Differences in functional connectivity based on rs-fMRI have been linked to cognition and behavior (Magnuson et al. 2015; Smith et al. 2015; Thompson, Magnuson, et al. 2013), as well as changes in vigilance level (Chang et al. 2013; Tagliazucchi and Laufs 2014; E. A. Allen et al. 2012). More importantly, these functional connectivity differences can discriminate between patient populations and healthy controls (Abbas, Bassil, et al., 2019; Du et al., 2018; Engels et al., 2018; Long et al., 2020; Miller et al., 2016; Rombouts et al., 2005; Sakoğlu et al., 2010; Sorg et al., 2007). However, because the BOLD signal is only loosely and indirectly linked to the underlying neural activity, findings regarding such differences in functional connectivity can often be difficult to interpret. Over the last 20 years, rodent models have proven instrumental in understanding the neuronal and neurophysiological basis of spontaneous BOLD signal fluctuations and the consequent intrinsic functional connectivity (Pais-Roldán et al., 2021). The number of rs-fMRI studies in rodents increases each year (Figure 1), and such studies will continue to play a key role in interpreting the alterations of functional connectivity that are observed across conditions and groups.

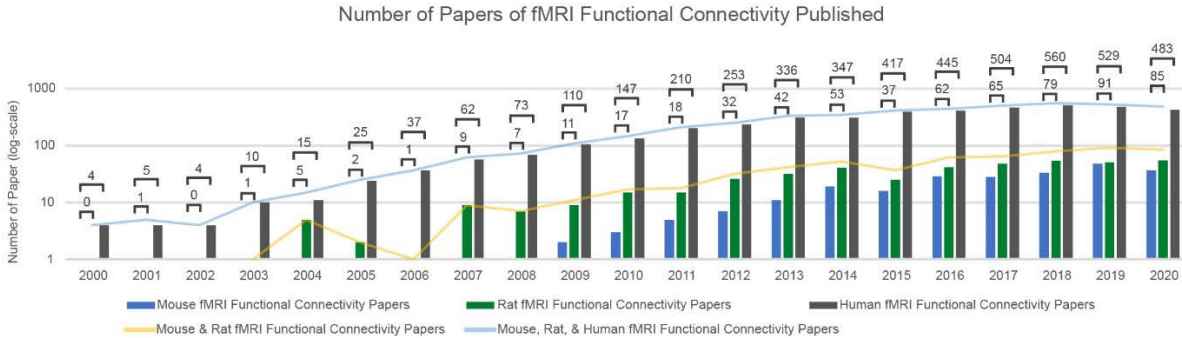

**Figure 1.** *The number of papers of fMRI Functional Connectivity Papers Published over the last 20 years. Utilizing the Web of Science* (Clarivate, 2017) *to search the last 20 years of published articles, an increasing and consistent trend of rs-fMRI studies for both human and rodent fMRI data is prominent. Each frequency search of published articles over a given year is scoped over publication titles, abstracts, and keywords. The keywords included in every search were: fMRI and functional connectivity. Each search was then delineated by year and the combinatorics of mouse, rat, and human (excluding "non-human" to avoid primate-specific studies). The y-axis is in a logarithmic scale to best characterize the increase of rodent papers over time.*

Much of the same general neural architecture is present in both humans and rodents, which accounts for the preponderance of rodent work in neuroscience. At the same time, differences in brain size, relative volumes of brain areas, and the presence or absence of cortical folding present obvious differences across species. In this manuscript, we aim to provide a comprehensive overview of features extracted from rs-fMRI studies in humans and rodents (rats and mice), with the goal of identifying similarities and differences that are important for the translation of knowledge gained in rodents to studies in humans. Resting-state networks are the most commonly reported features for both humans and animals, but we intend to go beyond basic functional connectivity analysis to report commonalities in network dynamics, community structure, and application to disease characterization. Before delving into the main aim, however, we will first touch upon the image acquisition parameters as well as the relative influences of the well-known fMRI signal confounds, such as motion and physiological noise, in both species. We hope that this manuscript will highlight areas for further investigation into important species-specific features and identify commonalities across species that increase confidence in investigations of animal models of brain disorders.

## II. Motivation for rs-fMRI studies in the rodent

The power of rs-fMRI lies in its ability to acquire information about activity noninvasively throughout the whole brain, making it suitable for use in healthy human subjects. This noninvasive characterization of brain function can be matched with invasive or time-consuming manipulations in rodents to provide insight into the neurophysiological processes that give rise to the BOLD fluctuations and create a better framework for the interpretation of rs-fMRI studies in humans. As a result, rodent studies have been invaluable for understanding the neural basis of rs-fMRI. Noninvasive techniques such as electroencephalography (EEG), electrocorticography (ECoG), or magnetoencephalography (MEG) that are suitable for human studies can measure extracranial electrical or magnetic fields resulting from neural activity but have limitations in terms of sensitivity, localization, and resolution. In rodents, however, the BOLD signal can be directly compared to valuable invasive methods that are not feasible in humans, including concurrent microelectrode recordings (Chen et al., 2019) or intracranial optical images of neural activity (Hewson-Stoate et al., 2005; H Lu et al., 2007; Ying Ma et al., 2016; Mateo et al., 2017). These studies have definitively demonstrated that the spontaneous BOLD fluctuations used to map functional connectivity are related to coordinated neural activity, placing rs-fMRI in humans on firm ground (Winder et al., 2017). As the field of rs-fMRI moved towards time-varying analysis, these same tools were crucial in demonstrating that while some of the variances are due to properties of the signal and analysis, some of them are a result of true variability in neural activity (Chan et al., 2015; S D Keilholz et al., 2013; Garth John Thompson, Merritt, et al., 2013).

Tools for genetic manipulation are available and particularly well-developed in rodents, making analysis possible to understand the genetic underpinnings of the structural and functional architecture of the brain. Two of the most prominent tools involving genetic manipulation, optogenetics and chemogenetics, use light and pharmacological agents respectively, to manipulate the activity of neurons. The combination of either of these techniques with fMRI allows for the stimulation or suppression of a specific cell type or pathway while simultaneously imaging the whole-brain functional response to the manipulation. For example, optogenetic-fMRI (ofMRI) has been used to examine the BOLD response to the activation of specific cellular populations (J. H. Lee et al., 2010; Weitz et al., 2015), and optogenetic and chemogenetic manipulations and their findings are beginning to provide insight into the complex interaction of neuromodulatory systems and localized activity (Albaugh et al., 2016; Weitz et al., 2019; Zerbi et al., 2015).

Rodent models could open the possibility of probing alterations that are observed in patient populations. For neurological disorders in particular, due to the origins of the BOLD signal, it is often difficult to tell if a change in functional connectivity arises from an alteration in the structural connections between areas, an alteration in the activity of the areas, or an alteration in the vasculature that affects neurovascular coupling. If, however, an animal model of the disorder presents the same alteration in functional connectivity, its underlying causes can be investigated using a variety of imaging, recording, and histological techniques. The mouse model, in particular, due to its universally-available transgenic animals, has served as a major tool for exploring pathogenic mechanisms of a variety of brain diseases (Denic, Macura, et al., 2011; Whitesell et al., 2019). A diagram of the positive feedback cycle of the cross-species study is illustrated in Figure 2.

One of the important motivations of rodent fMRI involvement could be taking advantage of minimum motion level procedures and fine controlling physiological conditions with ventilation in anesthesia and muscle relaxation in conventional experimental rodents. BOLD fMRI quality has been typically related to head motion levels that may introduce uncertainty in both image artifacts and brain state in physiological state variation. For instance, rat fMRI in ventilation may have minimized head motion, roughly multiple folders or even one order smaller than human data in translation and rotation of all 3 directions, especially respiration-related direction, demonstrated in Figure S1 (A) and (B).

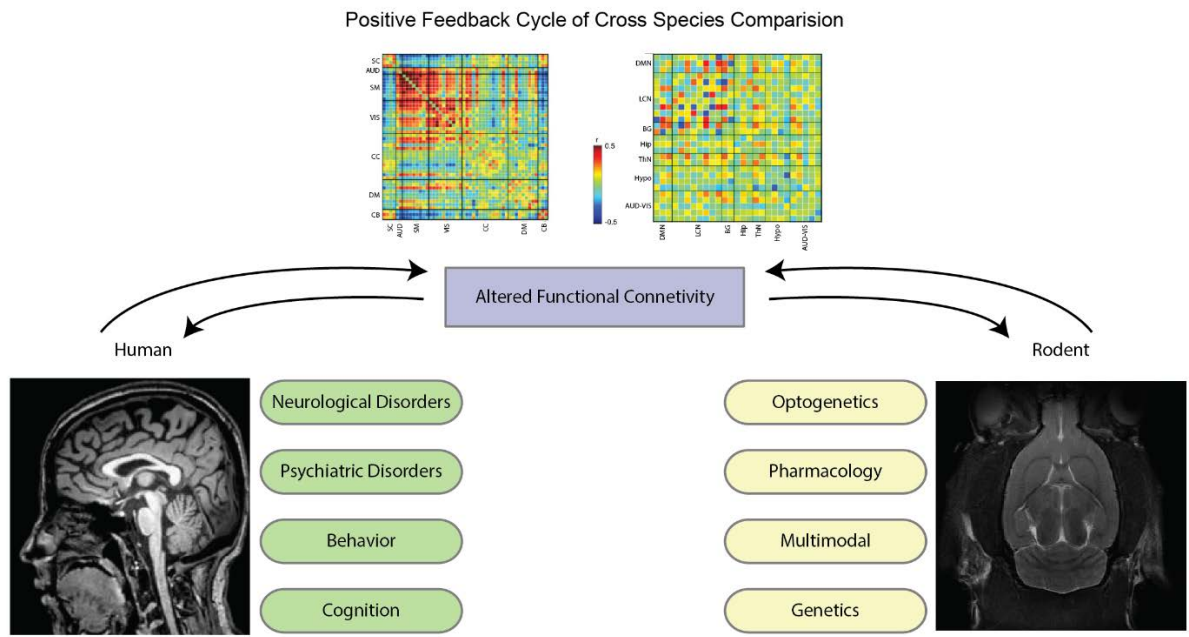

**Figure 2.** *Illustrative flow diagram of positive feedback cycle of cross-species comparison. With the advancements in imaging methods in the last decade, functional connectivity studies in both humans and rodents create a positive feedback cycle to best inform each other. For humans, neurological disorders (e.g., Alzheimer's Disease, ADHD, Epilepsy, etc.), psychiatric disorders (anxiety, depression, PTSD, etc.), behavior (task studies), and cognition are at the forefront of functional connectivity analysis. For rodents, utilizing tools such as optogenetics, pharmacology, genetic modification, and multimodal imaging informs novel insight into the brain*

*that warrants further investigation in human data. Throughout this paper, these contexts and tools will be explained in further detail, each of which captures specific data to better inform how the brain behaves in normal and perturbed conditions. By utilizing these methods, we are able to cycle back and forth between study paradigms to glean meaningful results. Functional connectivity matrices are excerpted from (Tsurugizawa & Yoshimaru, 2021) for mice (left) and from (Allen et al., 2012) for human brains. The human MRI image (left) comes from the Human Connectome Project (Van Essen et al., 2012), and the mouse MRI image comes internally from the Keilholz MIND Lab.*

## III. Acquisition parameters, conditions, and confounds in rodents compared to humans

*Size of anatomical features.* The most obvious difference between the brains of mice, rats, and humans is the size (Table 1). From a purely anatomical point of view, there is a clear size difference between species; 415 mm$^3$ volume of the mouse brain (Kovacĕvic et al., 2005) compared to the 1765 mm$^3$ volume of the rat brain (Welniak-Kaminska et al., 2019) compared to the 1200 cm$^3$ of the human brain (Yu et al., 2013). Ratiometrically, an isotropic voxel size of 1 mm in the human brain is equivalent to an isotropic voxel size of 114 um in the rat brain or an isotropic voxel size of 70 um in the mouse brain. This level of spatial resolution is challenging but achievable because the small size of the rodents makes it feasible to construct high field scanners (7T and up) that have higher SNR compared to fields typically used in humans (typically 3T, moving to 7T). For example, the SNR for human brain EPI data is typically 100 at 3T (B. T. Yeo et al., 2011) and it has been measured at 146+/-11 at 7T (Van Der Zwaag et al., 2012); in comparison, SNR for rodent brain EPI data up to 237.6 +/- 8.9 for multiple-slice EPI could be achieved at 7T in rats (Tambalo et al., 2019) and up to around 293.3 ± 37 in anesthetized mice under a combination of medetomidine-isoflurane and imaged at 9.4T with cryoprobes (Grandjean et al., 2020).

For functional images, the resolution is typically ~2 mm in humans, 200-400 µm in mice, and 300-500 µm in rats. For comparison, ocular dominance columns in humans were measured to have a mean diameter of 863 microns, and whisker barrel columns in mice and rats (which are essentially anatomically identical as reported by (Petersen, 2007)) are approximately 150 microns in diameter (Adams et al., 2007; Woolsey & Van der Loos, 1970). Cortical thickness is more consistent across species, ~0.9mm in mice, ~1-2mm in rats, and ~1-4.5 mm in humans (Fischl & Dale, 2000a; Lerch et al., 2008; Yu et al., 2013). These features can be found in Table 1. Unlike the gyrate cortex of humans, the cortex of rats and mice is smooth rather than folded, which tends to reduce the mixing of white and grey matter within a single rs-fMRI voxel (Ventura-Antunes et al., 2013).

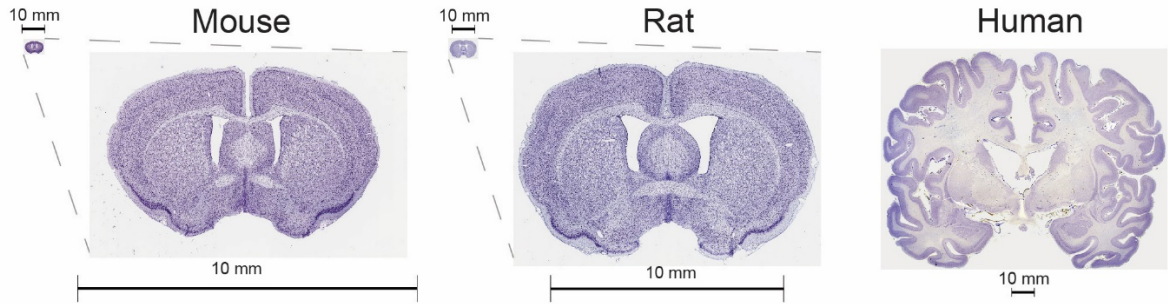

| | | **Mouse** | **Rat** | **Human** |
|---|---|---|---|---|
| Anatomy | Brain Volume | 415 mm3 (Kovacčvic et al., 2005) | 1765 mm3 (Welniak-Kaminska et al., 2019) | 1200 cm3 (Yu et al., 2013) |
| | Gyrification | Flat (Ventura-Antunes et al., 2013) | Flat (Ventura-Antunes et al., 2013) | Folded (Ventura-Antunes et al., 2013) |
| | Cortical Thickness | <1 mm (Lerch et al., 2008) | 1 - 2 mm (Vetreno et al., 2017) | 1 - 4.5 mm (Fischl & Dale, 2000) |
| MRI Parameters | Resolution | 20 - 50 μm (Turnbull & Mori, 2007) | 90 - 150 μm (Barrière et al., 2019) | 0.1 - 3 mm (Edlow et al., 2019; Fischl & Dale, 2000) |
| | TR (seconds) | 0.15-3 (Gilbert et al., 2019; Turnbull & Mori, 2007) | 0.15-3 (Gilbert et al., 2019) | <0.25 − 5 (K. H. Chuang & Chen, 2001; Sahib et al., 2018; Van Dijk et al., 2010) |
| | Resting-State Scan Time (minutes) | ~10-20 (Bauer et al. 2017; Belloy et al. 2018) | ~10 (Christiaen et al., 2019) | ~5-15 (Elliott et al., 2019) |

**Table 1.** *Overview and summary of brain features between mice, rats, and humans. The first row demonstrates the relative sizing of each species with a coronal Nissl-stained slicing from the BrainMaps initiative (Mikula et al., 2007). Each subsequent row compares common features of both anatomy and MRI parameters associated with mice, rats, and humans concerning functional connectivity studies.*

*Acquisition parameters.* Gradient-echo echo-planar imaging (EPI) is the most widely used acquisition technique for both human and rodent studies. Echo times must be shorter in rodents due to the shorter T2* at high magnetic field strength. The small size of the rodent brain combined with the high field and proximity of susceptibility gradients can also cause substantial distortion and signal dropout. On the positive side, stronger gradients can be used to provide the necessary high spatial resolution. Rodent data is frequently non-isotropic, with an in-plane resolution of 0.2-0.5 mm and 0.4-1.0 mm slice thickness (Becq et al., 2020; M. E. Belloy et al., 2021; Liska et al., 2015; Schwarz et al., 2013a; Tsai et al., 2020) while isotropic rodent data has ~0.5mm$^3$ voxels (Kundu et al., 2014) compared to about 2-3 mm$^3$ voxels for isotropic human data (Franzmeier et al., 2017). Repetition rates (TRs) in these studies are typically 1-3 s for both humans and rodents. More recently, advances in ultrafast fMRI have pushed the TR to subsecond for both humans (Feinberg et al., 2010; Sahib et al., 2018) and rodents (Gilbert et al., 2019; H.-L. Lee et al., 2019). Researchers often take advantage of the stability of anesthetized animals to acquire multiple short scans or a single set of data for a longer time (Pan et al. 2015). These parameters can be found in Table 1.

*Anesthesia.* A critical difference between typical rs-fMRI studies in humans and rodents is that anesthesia is typically used in rodents to prevent motion and minimize stress. For task-related fMRI, this has limited rodent studies to simple sensory stimuli, since studies of emotion, behavior, and/or task are not possible in animal models under anesthesia. This is less of an issue for rs-fMRI, where no stimulus is required. Notably, the anesthesia condition may be considered as a simplified but unique brain state for investigating brain intrinsic network activity with minimum peripheral inputs and outputs in addition to the benefits of fine physiological control on experimental rodents. However, anesthetics alter cerebral blood flow and blood volume, which impact the BOLD signal, along with the neural activity. For example, medetomidine is a noradrenergic antagonist that can produce undesired effects on whole-brain fMRI BOLD signals (Valerio Zerbi et al., 2019). Different anesthetics may potentially affect neural activity and neurovascular coupling in different ways (W. J. Pan et al., 2013). For example, isoflurane increases cerebral blood flow (CBF), which leads to a lower CBF and BOLD response to hypercapnia than in awake rats (Sicard et al., 2003). In contrast, CBF decreases in rats given dexmedetomidine (Ganjoo et al., 1998). The frequency of spontaneous BOLD fluctuations is increased in the mice under a combination of isoflurane and medetomidine (Grandjean, Preti, et al., 2017) which is different from 1/f infraslow spectra usually observed with light isoflurane and halothane (Gutierrez-Barragan et al., 2019) or in awake mice (Gutierrez-Barragan et al., 2022) and humans (Fox et al., 2007).

Specifically, for rs-fMRI, networks are more highly localized under light isoflurane anesthesia and become less spatially specific under deep isoflurane anesthesia (e.g., 1.8 % isoflurane), and much of the specificity under deep isoflurane anesthesia was recovered after global signal regression (Xiao Liu et al., 2013). While this may indicate that the procedure of global signal regression could potentially help control the widespread vasodilation that a high concentration of isoflurane induces, the effects of this procedure remained to be highly controversial (Murphy & Fox, 2017). In mice, different anesthetic agents result in different levels of interhemispheric connectivity (Jonckers et al., 2014). Many of the macroscale features of functional connectivity are observed under different anesthetics in rats, but different anesthetics have specific effects on particular connection types (Becq et al., 2020).

In addition, rs-fMRI in awake rodents has been explored with a measure of reduced stress levels, despite the challenges involved, with a number of groups successfully conducting rs-fMRI studies in unanesthetized rats or mice (Jonckers et al., 2014; Xingdan Liu & Huang, 2020; Yuncong Ma et al., 2020; Zilu Ma et al., 2018; Stenroos et al., 2018). Particularly, in an awake mouse study (Gutierrez-Barragan et al., 2022), though networks in the awake state were appeared to be anatomically conserved to the ones under anesthesia, evidence shows that there are network-specific regional anticorrelation pertaining to arousal-related basal forebrain areas and increased between-network crosstalk. This evidence is consistent with prior human anesthetic consciousness studies where transient anticorrelations between visual-auditory networks and the default mode network are observed by state-dependent interactions involving the basal-forebrain arousal system (Demertzi et al., 2019) and stereotypic temporal transitions between networks (Z. Huang et al., 2020). However, the process remains time-consuming and resource-intensive, and the relatively higher motion and certain stress levels limit the widespread use of unanesthetized rodents. This developing line of work in awake and anesthetized rodents remains an open question in fMRI research for best practices and universally accepted protocols.

*Motion.* Motion varies greatly in rodent rs-fMRI depending on whether awake or anesthetized animals are used. In anesthetized and ventilated rodents, motion is minimal compared to humans as demonstrated in Figure S1. Rodents are secured with a stereotaxic head holder and the anesthesia prevents voluntary motion. Sometimes paralytic agents are administered if the rodents are mechanically ventilated. For awake animals, the situation is very different. While healthy human subjects are generally able to remain steady during image acquisition with clear verbal communication, animal models require acclimation and head fixation methods (commonly implanted fixation points) to ensure firmly holding in place. Rodents may require only a few days or up to weeks of acclimation (King et al., 2005). Typical levels of head motion (average relative motion to voxel size) range from 2-3 % in awake animals, compared with 4-10% in humans to almost no head motion in anesthetized animals (Paasonen et al., 2018). Similar correction techniques can be applied in both animals and humans for displacing correction due to thermal drift in scanning. However, the anesthetized animals have been merely exhibiting spike head motion to be discarded (Paasonen et al., 2018). Additionally, some studies have shown that typical motion and nuisance correction for rodent MRI studies may not be suitable for proper analysis but rather for removing high motion rodents from a dataset as the correction did not significantly improve the signal fidelity (Chuang et al. 2018).

*Respiration and cardiac pulsation.* Respiratory and cardiac rates are much faster for rodents than for humans. Rates vary across individuals and strains but are on the order of 85 breaths per minute/300 beats per minute for rats (Carnevali et al., 2013) and 160 breaths per minute/630 beats per minute for mice (*MGI-Mouse Facts*, n.d.) in unanesthetized animals. The high frequencies of physiological cycles mean that the image sampling rate must be very high to avoid aliasing. In rats, a rs-fMRI study with a TR of 100 ms that could directly resolve the primary peaks associated with cardiac pulsation and respiration found that cardiac pulsation was most prominent near the base of the brain, while respiration contributions were most prominent near the sagittal sinus (Williams et al., 2010). As in humans, most rodent rs-fMRI studies use longer TRs (1-2 s) to allow whole-brain coverage, which means that respiratory and cardiac contributions can alias into the frequencies of interest. Unlike in human rs-fMRI studies, where variations in heart rate and respiratory rate are known to contribute to the BOLD fluctuations along with the primary frequency components, little has been done to examine similar variability in rodents. Indeed, respiratory variation is minimal when mechanical ventilation is applied. In line with the observation that widespread signal changes occur during varied breathing patterns in humans (Power et al., 2018), artifacts are increased in spatial extent in spontaneously breathing animals compared to ventilated animals (Pais-Roldán et al., 2018). Some part of this artifact can arise from the motion of the chest in the magnetic field, which induces more artifacts at high magnetic field strengths when the chest is in close proximity to the coil, as it is in small animal studies (Kalthoff et al., 2011).

## IV. Time-averaged features of rs-fMRI across species

*Homologous functional networks across species.* Functional connectivity refers to coherent changes in the BOLD fluctuations from different areas of the brain, often captured as correlation values. Regions that then show a high statistical association are referred to as a resting-state network (RSN). In humans, there are a variety of nomenclatures that are used to describe RSNs. A broadly used set of RSNs defined on human cortical regions was determined by (B. T. Yeo et al., 2011), which includes the default, the somatomotor, the visual, frontoparietal, dorsal attention, salience/ventral attention, and limbic networks. In rodent studies, several functional networks have also been commonly identified from resting-state fMRI data in anesthetized mice (Grandjean et al., 2020; Grandjean, Giulia Preti, et al., 2017; Stafford et al., 2014), anesthetized rats (Liang et al., 2012), awake rats (Liang, King, and Zhang 2011; Zhiwei Ma et al. 2018), and awake mice (Gutierrez-Barragan et al., 2022). Many of these brain networks are homologous to those of primates, for example, the default mode network (Grandjean et al., 2020; Grandjean, Zerbi, et al., 2017; Hsu et al., 2016; Stafford et al., 2014), somatosensory network (Grandjean et al., 2020; Grandjean, Zerbi, et al., 2017), and subcortical system (Grandjean et al., 2020), though it is well noted that the anesthetization causes a profound impact on the RSNs being identified (Liang et al., 2011, 2012; Zilu Ma et al., 2018).

Table 2 lists the homologous RSNs across humans, rats, and mice. For the sake of comparing the macroscale networks observed in the human brain to those observed in the rodent brain, we have chosen to use the seven network parcellation from (B. T. Yeo et al., 2011), which captures the large-scale functional architecture of the brain without being too granular for high-level comparisons across species. The seven Yeo RSNs detected in cortical areas are used as the reference, and the homologous functional networks in rats and mice brains and the corresponding major anatomical regions are described. For rats, the cortical RSNs defined in (Zhiwei Ma et al. 2018) are primarily used except for cingulate and prefrontal areas, which can be further divided into the limbic network (Barrière et al., 2019) and areas in default mode network (DMN) (Hsu et al., 2016; H Lu et al., 2012). Additionally, the rat salience network converged by functional and structural connectivity as is reported in (Tsai et al., 2020) is also included. For mice RSNs, brain regions that are commonly found in (Mandino et al., 2021; Sforazzini et al., 2014) for DMN and for salience network are included in the table. For other areas, the homologous networks defined in (Zerbi et al., 2015) are used. Finally, the lateral cortical network (LCN), which is postulated to perform the central executive function in rodents, is also included for rats (Schwarz et al., 2013) and for mice (Mandino et al., 2021). Note that the LCN is essentially the same as the somatosensory/motor areas for rodents (Grandjean et al., 2020). Yet, due to the multi-functions of this network, in contrast to the somatomotor and frontoparietal networks in humans, it is listed as the somatosensory/motor when it's compared to the somatomotor in humans, and it is listed as the LCN when compared to the frontoparietal in humans in Table 2.

While the RSNs described in (B. T. Yeo et al., 2011) only focus on the cortex, large portions of RSNs detected in rodent brains also include other brain areas. For example, in the pioneering work of functional parcellation of awake rat brain, the reported RSNs also include the brainstem, midbrain, thalamus/hypothalamus, amygdala, striatum, and hippocampus/retro hippocampus (Zhiwei Ma et al. 2018). Similarly, while cortical RSNs discovered in (Zerbi et al., 2015) only include somatosensory, sensory, and olfactory processing networks (Zerbi et al., 2015), the limbic system networks (shown in Table 2), as well as the basal ganglia and cerebellar networks (that are not listed in the table) are all non-cortical networks. The cross-species comparison of each homologous network is discussed below.

As shown in Table 2, visual and somatomotor networks can be robustly detected in humans, rats, and mice by using a whole-brain functional analysis. The visual networks in humans and rodents, despite the overall differences (Jonckers et al., 2011; Katzner & Weigelt, 2013), share structural and functional principles, which allow for investigation of visual processing from the cellular level (using rodents models) to the macroscopic level (in human studies) (Katzner & Weigelt, 2013). For the whole somatomotor cluster, separate networks of somatosensory, motor, auditory, and olfactory were discerned in the functional network analysis of rodents which encompass the anatomical regions that are named after (Zhiwei Ma et al. 2018; Zerbi et al. 2015). However, the somatosensory, motor, and auditory cortices were all grouped into one somatomotor network in human studies (B. T. Yeo et al., 2011). Similar to humans, the somatomotor network in rodents is responsible for movement and sensory. However, different from humans, the somatomotor network in rodents is also postulated to be responsible for central executive control (Gozzi & Schwarz, 2015; Grandjean et al., 2020), which is a function of the human frontoparietal network instead. This function is described in the latter paragraph with the human frontoparietal network.

| Human networks (Yeo et al. 2011) | anatomical regions | Rat networks | anatomical regions | Mouse networks | anatomical regions |
|---|---|---|---|---|---|
| **Visual** | Visual cortex | **Visual areas** (Zhiwei Ma et al. 2018) | Rostral and caudal part of visual cortex, parietal cortex, and retrosplenial cortex | **Sensory, Visual** (Zerbi et al. 2015) | Visual cortex and retrosplenial dysgranular |
| **Somatomotor** | Motor cortex | **Motor** (Zhiwei Ma et al. 2018) | Primary and secondary motor cortex, mammillary nucleus, ventral hypothalamus | **Sensory, Motor** (Zerbi et al. 2015) | Motor cortex |
| | Auditory cortex | **Auditory** (Zhiwei Ma et al. 2018) | Auditory cortex | **Sensory, Auditory** (Zerbi et al. 2015) | Dorsal and ventral auditory cortex |
| | Somatosensory cortex posterior insular cortices | **Somatosensory** (Zhiwei Ma et al. 2018) | Primary and secondary somatosensory cortex, posterior part of the insular cortex | **Somatosensory** (Zerbi et al. 2015) | Upper lip region, barrel field, hindlimb region, and forelimb region in primary somatosensory cortex |
| **Limbic** | Orbital frontal cortex | **Olfactory** (Zhiwei Ma et al. 2018) | Piriform cortex, anterior olfactory nucleus, and olfactory tubercle | **Sensory, Olfactory** (Zerbi et al. 2015) | Piriform cortex, medial orbital cortex, and glomerular layer of the olfactory bulb |
| | Temporal pole | **Limbic** (Barrière et al., 2019) | Prelimbic cortex, prelimbic/infralimbic | **Limbic system (non-cortical)** (Zerbi et al. 2015) | Cingulate cortex area 1, 2, and retrosplenial cortex Ventral and dorsal hippocampus, amygdala |
| **Default** | Precuneus posterior cingulate cortex/retrosplenial cortex | **Default mode network (DMN)** (Hsu et al., | Cingulate cortex, retrosplenial cortex | **DMN** (Mandino et al., 2021; Sforazzini et al., 2014) | Cingulate cortex, retrosplenial cortex |
| | Prefrontal cortex | 2016; Hanbing Lu et al., 2012) | Orbital cortex, prelimbic cortex, association visual cortex | | Prefrontal and orbito-frontal cortex, prelimbic cortex |
| | Parietal and temporal lobes | | Posterior parietal cortex, auditory/temporal association cortex | | Temporal cortex |
| | Parahippocampal cortex | | Hippocampus (CA1) | | Ventral-hippocampus, dorsal striatum, dorsolateral nucleus of the thalamus |
| **Frontoparietal** | Parietal and temporal lobes<br>Dorsal, lateral, and ventral prefrontal cortex<br>Orbital frontal cortex<br>Precuneus, cingulate, medial posterior prefrontal cortex | **Lateral cortical network** (Schwarz et al., 2013) | Anterior secondary motor cortex, secondary sensory, insula | **Lateral cortical network** (Mandino et al., 2021) | Primary motor, primary somatosensory, lateral striatum, ventroposterior nucleus of the thalamus |
| **Dorsal attention** | Posterior, frontal eye fields, precentral ventral | | | | |
| **Salience/ventral attention** | Parietal operculum, temporal occipital, frontal operculum insula, lateral prefrontal cortex, medial nodes | **Salience*** (Tsai et al., 2020) | Anterior portion of the agranular insular, frontal cortices including the anterior cingulate cortex | **Salience*** (Mandino et al., 2021; Sforazzini et al., 2014) | Anterior insular, dorsal anterior cingulate, ventral striatum/nucleus accumbens |

**Table 2:** *Homologous functional networks across mice, rats, and humans. * indicates that the network that is not exactly homologous but shares similar functions in some partial regions.*

The olfactory network in humans was not discerned as a separate network in the whole brain RSN analysis as was revealed in rodents, although a recent study was able to detect the human olfactory network using functional connectome analysis (Arnold et al., 2020). The olfactory network in rodents typically includes the main and accessory olfactory systems, which are responsible for the sense of smell and pheromone-based communication, respectively (Huilgol & Tole, 2016). For humans, the olfactory network serves more heterogeneous functions, including not only olfactory sensory perception but also multiple non-sensory functions such as emotion, neuroendocrine, and homeostasis. Anatomically, the human olfactory network is widely distributed over cortical and subcortical areas, which can be further decomposed into limbic, frontal, and sensory systems (Huilgol & Tole, 2016).

The limbic network, acting as one of the most complicated systems in the brain, is involved in various brain functions, including homeostasis, memory, emotions, olfaction, and many more (Moini & Piran, 2020). Note that the limbic network coverage listed in Table 2 only includes the cortical structures (B. T. Yeo et al., 2011). Yet, the human limbic network indeed extends to subcortical and interbrain regions—including the amygdala, hippocampus, septal nuclei, and hypothalamus (Rajmohan et al., 2007; Sokolowski & Corbin, 2012). Similar limbic structures and hence similar functions have been discovered in humans and rodents, despite the enlarged olfactory bulbs and the presence of the accessory olfactory bulbs in rodents (Sokolowski & Corbin, 2012). Due to the similar processing of emotions and other social cues in the limbic network for both rodents and humans, limbic responses to social cues have been increasingly studied in rodent models, which were then translated to humans (Sokolowski & Corbin, 2012).

The default network or default mode network (DMN) is a network suppressed during tasks and activated during resting-state, such as during mind-wandering; it has also been found to be active during remembering, imaging the future, and making social inferences (Buckner, 2013; Buckner & DiNicola, 2019). A key feature of the DMN revealed by functional analysis is that it exhibits a pattern of anticorrelation with the parietal/fronto-cortical areas (namely the "task-positive network" as described in the next paragraph), which are supposed to be more active during tasks (M D Fox et al., 2005). Hence, DMN is sometimes referred to as the "task-negative work" (M D Fox et al., 2005). A homologous DMN or DMN-like network has been increasingly reported in both rats (Liang et al., 2012; H Lu et al., 2012; Upadhyay et al., 2011) and mice (Grandjean et al., 2020; Sforazzini et al., 2014; Whitesell et al., 2021; Zerbi et al., 2015), which exhibit a similar robust anti-correlation pattern with the parietal/frontal areas (Gozzi & Schwarz, 2015) and a passive role during tasks (Li et al., 2015; Schwarz et al., 2013). Additionally, similar within-network connectivity patterns were observed across rodents and humans (Gozzi & Schwarz, 2015). The anatomical coverage of the DMN was distributed along the medial and lateral parietal, medial prefrontal, and medial and lateral temporal cortices across rodents and humans (Raichle, 2015), with the critical components including the parietal cortex, the orbitofrontal cortex, and the cingulate areas (Stafford et al., 2014), as well as the temporal association areas (Barrière et al., 2019; Grandjean et al., 2020; Schaefer et al., 2018). Studies have suggested that the human cortical DMN anatomy may be compared with the rodent DMN from the molecular, anatomical, and all the way up to the functional level (Coletta et al., 2020; Whitesell et al., 2021). Leveraging the novel neuroanatomical technology such as viral tracing and whole-brain imaging, Whitesell et al. (2021) reveal the mice's DMN anatomical structure in high-resolution and identify cell type correlates of this network. In particular, revealed by both fMRI voxel-wise analysis and viral tracing, cortical DMN regions in mice appear to be preferentially interconnected. In addition, excitatory neuron classes in different cortical layers project to the mice DMN in different patterns. For example, Layer 2/3 DMN neurons project mostly in the DMN whereas layer 5 neurons project in and out (Whitesell et al., 2021). The homologous default networks are shown in Figure 3. Recent studies in both rodents and humans have further suggested that the DMN comprises multiple distinct but interwoven networks rather than a single network (Buckner & DiNicola, 2019). Moreover, in humans, the aberrant connections in the DMN or/and in its interplay with other networks have been found to constitute the "connectopathy" in many neurologic conditions such as Attention-deficit/hyperactivity disorder (ADHD) (Lucina Q. Uddin et al., 2008), Autism spectrum disorder (ASD) (L Q Uddin et al., 2019), Alzheimer's disease and ageing (Toga & Thompson, 2014). Notably, a similar "connectopathy" involving aberrant rodent's DMN was also discovered in recent studies using transgenetic mice for Autism Spectrum Disorder (Pagani et al., 2021) and Alzheimer's disease (Adhikari, Belloy, Van der Linden, Keliris, & Verhoye, 2021). Despite these similarities across rodents and humans, discrepancies remain. As is reviewed in (Gozzi & Schwarz, 2015), cross-species differences in the DMN still exist both in functional and in neuroanatomical organizations. Notably, the precuneus, which acts as the most prominent hub in the human DMN, lacks a clear neuroanatomical equivalence in rodents, and its functional role might be relegated to the retrosplenial cortex in rodents (Vogt & Paxinos, 2012).

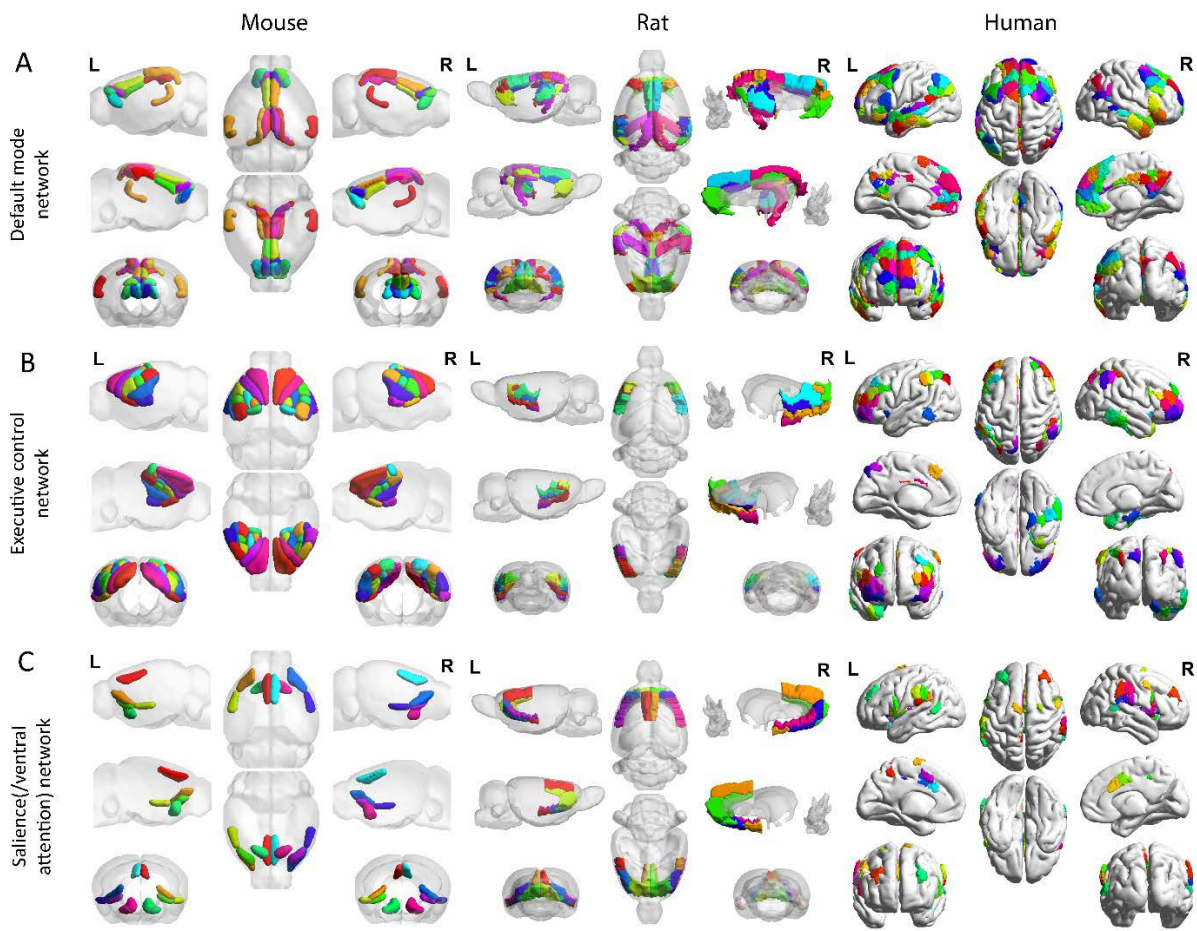

**Figure 3.** *Homologous default mode, executive control, and salience (/ventral attention) networks across mice, rats, and humans. For the DMN (A), the mouse (left) follows the region coverage exported in (Grandjean et al. 2020); the rat (middle) follows the region coverage exported in (Hsu et al. 2016); the human brain (right) shows the default network in (Yeo et al. 2011). For the executive control network (B), the mouse follows the region coverage of lateral cortical network exported in (Grandjean et al. 2020); the rat (middle) follows the lateral cortical network coverage exported in (Schwarz et al., 2013); the human brain (right) shows the frontoparietal network in (Schaefer et al., 2018; B. T. T. Yeo et al., 2011). For the salience(/ventral attention) network (C), the mouse (left) follows the salience network commonly found in (Mandino et al., 2021; Sforazzini et al., 2014); the rat (middle) follows the salience network coverage converged by functional and anatomical connectivity from the ventral anterior insular division (Tsai et al., 2020); and the human brain (right) shows the salience/ventral attention network reported in (Schaefer et al., 2018; B. T. T. Yeo et al., 2011). All mouse (left), rat (middle), and human (right) networks are demonstrated on the Allen Institute for Brain Science mouse atlas (Lein et al. 2007), the SIGMA anatomical atlas (Barrière et al., 2019), and the Schaefer-Yeo atlas (Schaefer et al., 2018), respectively. All figures were created by BrainNet Viewer (Xia, Wang, and He 2013).*

The frontoparietal network in humans is also known as the central executive network, which supports the executive function and goal-oriented, cognitively demanding tasks (Gozzi and Schwarz 2015). Though it has been questioned whether rodents have an exact homologous frontoparietal network (Gozzi & Schwarz, 2015), areas with similar functions have been increasingly found in both mice (Harvey, Coen, and Tank 2012; Tombaz et al. 2020) and rats (Brunton, Botvinick, and Brody 2013). For example, the posterior parietal cortex (PPC), a central component in the human frontoparietal network, is involved in visual attention, working memory, spatial processing, and movement planning in humans, and it is involved in navigation, sensory processing and decision making in rodents (Harvey et al. 2012). In addition, the frontoparietal network together with the dorsal attention network are commonly considered as the 'task-positive network' in humans, due to their co-activation during many 'attention-demanding' tasks (Power et al., 2011; Spreng et al., 2013). Studies have revealed that a similar 'task-positive network' is conserved in rodent brains, which are primarily located in the somatomotor areas (Gozzi & Schwarz, 2015; Grandjean et al., 2020; Mandino et al., 2021). This network, namely the lateral cortical network (LCN), is postulated as the central executive network in rodents (Gozzi & Schwarz, 2015). The homologous central executive network is shown in Figure 3 (B).

The salience/ventral attention network encompasses the salience network and ventral attention network, two of which are found to be highly associated (Schaefer et al., 2018; B. T. T. Yeo et al., 2011) and are tied to multiple attention components including error monitoring, selective attention, and task switching (Michael D. Fox et al., 2006; V. Menon & Uddin, 2010). The salience network, in particular, is considered to segregate unexpected but salient stimuli (e.g., top signal), as well as to modulate the switch between the internally directed cognition (DMN) and the externally directed cognition (central executive network) (Menon and Uddin 2010). It has many overlapping functions with the ventral attention network which mainly enacts in the bottom-up attentional process (Hsu et al. 2020; Vossel, Geng, and Fink 2014). Converging evidence has revealed that the salience network is evolutionarily conserved across mice (Gozzi & Schwarz, 2015; Grandjean et al., 2020; Mandino et al., 2021), rats (Tsai et al., 2020), and humans (Gozzi & Schwarz, 2015;

Grandjean et al., 2020; Mandino et al., 2021; Tsai et al., 2020). Similar to humans, the salience network in rodents is anchored in the insula, and extends to the dorsal anterior cingulate cortex, and also includes other subareas such as the amygdala (Grandjean et al., 2020; Mandino et al., 2021; Tsai et al., 2020). The homologous functional salience network of mice (Mandino et al., 2021) and rats (Tsai et al., 2020) as well as the salience/ventral attention network (Schaefer et al., 2018) are shown in Figure 3(C). Note that though convergence between functional and anatomical connectivity in rat salience network was only reported in ventral anterior insular areas (Tsai et al., 2020), the other projections and connections, including ones between the amygdala and thalamus and the insular in rat salience network are well confirmed by anatomical tracers (Bota et al., 2015; Tsai et al., 2020). Similarly, while Table 2 and Figure 3(C) only report the salient network regions that are commonly found in (Mandino et al., 2021; Sforazzini et al., 2014), other important areas including the central and basolateral amygdala are also recently discovered (Mandino et al., 2021). The rat salience network demonstrates close correspondence to the mouse salient network (Mandino et al., 2021; Tsai et al., 2020). Additionally, as noted in (Mandino et al., 2021), DMN and salience network have overlaps in the anterior cingulate cortex, where the salience network is located more in the anterior part and DMN more is more in the posterior part.

Apart from the above increasingly discovered homologous networks, the existence of homologous RSNs involved in other higher cognitive functions in rodents is still left as an open question. In particular, the dorsal attention network and the ventral attention network (the other part of the salience/ventral attention network) have not yet been discovered in rodents (Vossel et al., 2014). These two attention networks in humans are collectively involved in the control of attention related to top-down goals and bottom-up sensory stimulation (Vossel et al., 2014). It has been hypothesized that they might simply not be present or be subserved by some brain regions in other large networks such as the DMN (Gozzi & Schwarz, 2015).

Cross-species studies may also demonstrate that there are inconsistent neurological patterns between species that may require different processing methods. For example, rats and mice exhibit cortical differences in rs-fMRI data where unilateral cortical networks exist for mice versus bilateral networks in rats (Jonckers et al., 2011). Additionally, there is evidence demonstrating functional connectivity differences between non-human primates and rodents within the medial frontal cortex and regions of the lateral frontal cortex (Schaeffer et al., 2020). For both humans and non-humans, the border between subregions of the cingulate cortex is drawn along the rostrocaudal axis; however, in mice and rats, this boundary has been routinely drawn perpendicularly to this axis (Heukelum et al., 2020). Further, there are differences anatomically between the types of cells, regional cell packing across species (Abreu et al., 2020), and specific wiring of various structures where species with fewer interhemispheric connections exhibit better intrahemispheric connectivity (Y. Assaf et al., 2020).

*Global signal.* The global signal is obtained by averaging the BOLD timecourses over the entire brain. Fluctuations in the global signal have been linked to physiological fluctuations, mainly respiratory effects, head motion, scanner-related effects, and vascular effects (Murphy & Fox, 2017; Power et al., 2017). Along with these nuisance variables, the global signal also contains contributions from widespread coherent neuronal activity, and animal studies have been crucial to understanding this aspect. In an early study, Schölvinck and colleagues showed that the rs-fMRI signal calculated over the entire cerebral cortex in monkeys is positively correlated with the spontaneous fluctuations in the local field potentials though recorded from a single cortical site (Scholvinck et al., 2010). More recent studies in monkeys have demonstrated that the activity of the basal forebrain is linked to arousal level (Xiao Liu, De Zwart, et al., 2018) and global signal (Turchi et al., 2018), showing that inactivation of the basal forebrain leads to increased global spontaneous fluctuations. The spatial pattern of correlation of global signal and global signal amplitude decrease in the same way in rats and humans with increasing levels of anesthesia (Tanabe et al., 2020). In awake rats, the peaks of the global signal correspond to the time of high activity in the sensorimotor cortex, hippocampus, medial thalamus, prefrontal cortex, and basal forebrain (Yuncong Ma et al., 2020). Neither respiration nor head motion was significantly different across times of high and low global signal. It is possible that contributions from motion and physiological noise are reduced in anesthetized animals, where variability is minimized, as compared to humans.

*Relationship to structural connectivity.* A number of properties of functional connectivity are common across species. In general, there is a substantial overlap of structural and functional connectivity in both rodents and humans. In the rat, a meta-analysis of tracing data combined with rs-fMRI found a Spearman rank-order correlation of 0.48 between the structural and functional connectivity matrices (Díaz-Parra et al., 2017). In the mouse, previous studies (Grandjean, Zerbi, et al., 2017; Stafford et al., 2014), that used parceled connectome analysis on by the Allen Brain Connectivity Atlas (where parcels were a large piece of predefined brain regions), reveal that cortical functional connectivity is related to monosynaptic connectivity derived from viral tracers (Stafford et al., 2014); and bilateral cortical areas such as somatosensory cortex typically exhibit close correspondence with the structural connectivity, while functional connectivity in the DMN is only partially explained by monosynaptic connections and little connection was observed between anterior and posterior areas (Grandjean, Zerbi, et al., 2017). Yet, a recent study (Coletta et al., 2020), using a fine-grained voxel-wise brain-wide analysis, detects a much closer structural-functional correspondence in DMN. Specifically, strong structural and rich connections are detected in the entire DMN, including between the anterior cingulate and retrosplenial areas. This is consistent with the previous human study using diffusion-based tractography, which identifies strong connections between anterior and posterior areas that follow the medial wall of the cortex (Honey et al., 2009). In mice, functional connectivity in subcortical regions was mostly mediated by polysynaptic connections. Little evidence of functional connectivity was found for thalamocortical circuits, despite the monosynaptic structural connections, possibly due to the use of anesthesia, as thalamocortical connectivity has been observed in humans (D. Zhang et al., 2010).

*Functional connectivity gradients.* Functional connectivity gradients, typically obtained using dimensionality reduction on voxel-wise functional connectivity matrices, identify spatial axes along which voxels exhibit similar functional connectivity profiles (Margulies et al., 2016). Functional gradients have been linked to neuronal micro-architectures, such as myelin density or gene expression across the cortical sheet (Burt et al., 2018; Fornito et al., 2019), and reflect fundamental properties of the brain's functional organization. Functional connectivity gradients have also been observed in the mouse and tied to similar properties (although the relationship to cytoarchitecture was not significant) (Huntenburg et al., 2021). Many aspects of the gradients observed in mice recapitulate findings in humans (Coletta et al., 2020; Fulcher et al., 2019). Functional connectivity gradients in the mouse brain used to assess the intrinsic functional organization of the cortex reveal that a prominent gradient reflects the spatial distance from the presumed origins of cortical evolution (Fulcher et al., 2019). The primary functional gradient in the anesthetized mouse brain has been found to span from lateral cortical motor-sensory areas towards more transmodal components (Fulcher et al., 2019), specifically the DMN (Coletta et al., 2020), and a second functional gradient extends across unimodal visual and auditory cortices up to primary motor-sensory areas (Coletta et al., 2020). This may be due to the effect of the specialization of sensory areas on the intrinsic functional organization of the mouse cortex. Similarly, in humans, the primary functional gradient spans the sensorimotor-to-transmodal axis (i.e., from SM to DMN), which is also viewed as the axis of information integration and abstraction (Huntenburg et al., 2021; Margulies et al., 2016), and a second gradient separates somatomotor and auditory cortex from visual cortex, which indicates the integration comes across distinct modalities (Margulies et al., 2016).

*Graph metrics.* Functional connectivity can also be described in terms of a graph, where nodes represent different parcels of the brain and edges represent the correlation between them. Graphical analytical tools can then be utilized to describe the community structure of functionally connected regions. For node determination, many brain atlases have been developed for brain network analysis. In mice studies, the anatomical atlas provided by Australian Mouse Brain Mapping Consortium and the Allen Mouse Brain Atlas has been widely used (Grandjean et al., 2020; Grandjean, Giulia Preti, et al., 2017; Stafford et al., 2014). For adult rat brains, many brain anatomical atlases have been developed and some have been employed in recent whole-brain network analysis, e.g., the Swanson atlas (Swanson, 2018), and the frontier SIGMA atlas that developed from multimodal imaging data (Barrière et al., 2019). A detailed comparison of these rat atlases is described in (Barrière et al., 2019). In addition to these anatomical parcels, brain-wide functional parcels were also recently derived from the resting-state fMRI data for anesthetized mice (Grandjean et al., 2020; Grandjean, Giulia Preti, et al., 2017), anesthetized rats (Barrière et al., 2019), and awake rats (Zhiwei Ma et al. 2018). These functional parcels, which often overlap with multiple neighboring parcels from the anatomical atlas, profile the functional network connectivity. In the recent rodents' study, different approaches have been used for determining the data-specific functional parcels. For example, independent component analysis-based (ICA-based) methods have been popularly used in the study of anesthetized (Barrière et al., 2019; Grandjean et al., 2020; Grandjean, Preti, et al., 2017) and awake (Liang et al., 2011) rodents. In the recent study of deriving brain-wide functional parcels for awake rats, k-mean clustering on dissecting the spatial similarity/dissimilarity between functional connectivity profiles was also used (Zhiwei Ma et al. 2018). In human studies, more parcellation schemes become available for identifying functional parcels, for example, the frontier brain cortical atlas developed from both resting-state and task fMRI data (Schaefer et al., 2018), the Gordon cortical atlas that was purely developed from resting-state fMRI data (Gordon et al., 2016), and the Brainnetome atlas developed from structural MRI, rs-fMRI, and diffusion MRI which covers both the cortical and subcortical regions (Fan et al., 2016). A detailed comparison between different human brain atlases was discussed in (Eickhoff et al., 2018). Note that because the use of predefined regional parcellation automatically assumes homogeneity of connections within every parcel, it may limit spatial resolution and potentially omit fine-grained network structure. The unprecedented work from Allen Institute for Brain Science (Knox et al., 2018) relaxes such an assumption by introducing a regional voxel-wise mathematical model for mouse connectome. This novel voxel model assumes smoothness across major brain divisions instead (Knox et al., 2018), which has been used by researchers to construct a voxel-level description of directed mouse connectome unconstrained by regional partitioning (Coletta et al., 2020).

To determine the brain organization, clustering or graph theory-based community detection methods are often used (Coletta et al., 2020; Gordon et al., 2016; Knox et al., 2018; Power et al., 2011; B. T. Yeo et al., 2011). In particular, these methods cluster the functional connectivity map computed by the Pearson correlation from the BOLD signals extracted from the anatomical or functional parcels into distinct non-overlapping functional networks. In human studies, the clustering algorithm developed in (Lashkari et al., 2010) has been used for identifying seven common functional networks in the brain cortex (B. T. Yeo et al., 2011). Similar networks were obtained (Gordon et al., 2016; Power et al., 2011) by using the Infomap graphical analytical algorithm (Rosvall & Bergstrom, 2008). In mice, leveraging the regional voxel-wise analysis and then performing the agglomerative hierarchical clustering procedure, four networks-including DMN, lateral cortical network, hippocampal system, and olfactory/basal forebrain areas-are robustly detected (Coletta et al., 2020).

Previous studies have noted rich-club organization of neural structure centrality and network identification for the human (Grayson et al., 2014; M P van den Heuvel & Sporns, 2011), and mouse brain (Coletta et al., 2020; de Reus & van den Heuvel, 2014), as well as for a variety of other animal models: macaque (Harriger et al., 2012); mouse lemurs (Garin et al., 2021); cat (Zamora-López, Zhou, and Kurths 2009, 2010, 2011; de Reus and van den Heuvel 2013); avian (Shanahan et al., 2013); and neural systems of nematodes (*C. elegans*) (Towlson et al., 2013). For example, highly connected areas known as hubs tend to be densely interconnected, forming a "rich club" of brain areas (Sporns et al. 2007; van den Heuvel et al. 2011). In humans, these areas are found in regions of the brain such as the precuneus, frontal cortex, posterior parietal cortex, and thalamus. A similar distribution of rich club nodes has been observed in rodents, for example in the cingulate cortex and other areas of the default mode network, and in the thalamus (Coletta et al., 2020; Liska et al., 2015; Zhiwei Ma & Zhang, 2018; van den Heuvel et al., 2016). The existence of such rich club nodes for human and mouse brains supports the opinion that the state or extent of anatomical interconnection among brain regions governs the functional connectivity in spontaneous brain activity. This similarity between functional networks of rodent and human brains allows for the determination of fundamental indicators in the topology of neural circuits that can be used to translate to clinical studies of brain networks during normal and abnormal development. Most of these rich hubs are shown to be strongly mutually interconnected. One limitation of such analysis is the mismodeling of long-distance projections of neurons within each hub or rich club denoting by van den Heuvel et al. to which the degree of centrality needs to be relaxed and varies from study to study (van den Heuvel et al., 2016).

## V. Time-varying features of rs-fMRI across species

In the last decade, researchers have increasingly turned to a time-varying assessment of functional connectivity, which provides information about the evolving state of the brain, rather than the time-averaged picture obtained when functional connectivity is calculated over the course of the whole scan (Calhoun et al., 2014; R M Hutchison et al., 2013; Karahanoğlu & Van De Ville, 2017; Lurie et al., 2020; Preti et al., 2017). These dynamic brain states have shown great promise for studying both healthy and disordered brains (Calhoun et al., 2014; R M Hutchison et al., 2013), yet questions and controversies remain (R M Hutchison et al., 2013; S. Keilholz et al., 2017; Lurie et al., 2020). In the meanwhile, animal models have played an essential role in distinguishing variance of interest from changes unrelated to neural activity. Here we review commonly reported findings across species, along with neuronal correlates where known.

*Windowed approaches and brain states.* Sliding window correlation is one of the most basic approaches of time-varying analysis and involves calculating the correlation between pairs of brain areas (i.e., functional connectivity) for a time window from a scan, rather than for the entire scan (Allen et al., 2012; R. Matthew Hutchison et al., 2013; Shakil et al., 2016). The window is then moved along the scan timecourse to give a timecourse of correlation. Sliding window correlation offers advantages over time-averaged functional connectivity for identifying group differences (Damaraju et al., 2014; S. S. Menon & Krishnamurthy, 2019). However, scans of longer duration or additional scanning sessions are required for performing robust sliding window correlation analysis (Hindriks et al., 2016) and faster dynamics may be undetectable using a windowed approach due to practical limits on the minimum window length (Leonardi & Van De Ville, 2015).

Early studies in rodents examined sliding window correlation between a few pairs of areas and were instrumental in showing that apparent dynamics could arise from the properties of the BOLD signal itself (Keilholz et al. 2013). Nevertheless, simultaneous microelectrode recording and rs-fMRI showed that sliding window correlation of the BOLD signal at least partially reflects changes in the correlation of bandlimited power of local field potentials across areas (Garth John Thompson, Merritt, et al., 2013).

In humans, sliding window correlation is typically applied to parcellated brain signals, with the resulting matrices of functional connectivity over the course of the scan then clustered to identify common brain states (Allen et al., 2012). This approach is increasingly applied in rodent models as well (Grandjean, Giulia Preti, et al., 2017). Grandjean et al. used an ICA parcellation and calculated sliding window correlation for 45s windows and found distinct states that varied in terms of network dominance and interaction between the DMN, lateral cortical areas, and striatal areas, among others. Another study used the Allen Mouse Brain parcellation and identified states that varied in terms of their occurrence rates over different anesthesia levels (Yuncong Ma, Hamilton, and Zhang 2017).

In humans, there is evidence that the relative occurrence of brain states based on sliding window correlation reflects the relative arousal of the subject (Haimovici et al. 2017; Allen et al. 2012; Chang et al. 2013). It is not clear how this effect might manifest in anesthetized or sedated animals that are maintained at a stable condition. It may be that the variability across states is less than would be observed in humans, which does appear to be the case for Allen et al compared to Yuncong Ma et al, where connectivity is relatively similar across states. However, using an ICA-based parcellation, Bukhari et al. observed greater variability across states, and the occurrence of some states was linked to the level of anesthesia (Bukhari et al., 2018).

*Quasi-periodic patterns.* Quasi-periodic patterns (QPPs) are spatiotemporal patterns of activity that occur repeatedly throughout the scan. A pattern-finding algorithm has been developed (Majeed et al., 2011) and popularly used (Abbas, Belloy, et al., 2019; M. E. Belloy et al., 2021; Belloy et al., 2018; Yousefi et al., 2018; Yousefi & Keilholz, 2021) to detect the most prominent QPPs that are recurring over the course of the whole scan. Yet, recent advances allow us to understand these patterns within this continuous process more thoroughly by identifying spatiotemporal trajectories (X. Zhang et al., 2021). QPPs were first reported in anesthetized rats, where they appeared as bilateral waves of activity propagating from lateral to medial areas (Majeed M; Keilholz, SD et al., 2009). Shortly thereafter, similar patterns were found in humans using a pattern-finding algorithm, where they involve cyclical activation and deactivation of the DMN and task-positive network (TPN) (Majeed et al., 2011). More recently, QPPs have also been observed in healthy mice (Belloy et al. 2018).

Multimodal studies in rodents have been key to developing a better understanding of the neurophysiology underlying QPPs. Early work showed that they are present in CBV-weighted data, as well as BOLD (M. Magnuson et al., 2010). Using simultaneous microelectrode recording and MRI, further work showed that QPPs have links to infraslow electrical activity (Thompson et al. 2014; Pan et al. 2013). QPPs continue to occur after acute disconnection of the corpus callosum but are often unilateral, suggesting that the white matter connections may be necessary for the coordination of the pattern across hemispheres (Magnuson et al. 2014).

QPPs are similar across species in that they appear to involve analogous areas. In humans, activity propagates from the somatomotor network to the default mode network (Yousefi & Keilholz, 2021).  In rats, activity propagates from lateral somatosensory areas to medial cingulate cortex (Majeed et al., 2009, 2011), and in a recent study using whole-brain imaging in mice, QPPs involved anticorrelation between DMN areas and TPN areas, reminiscent of findings in humans (Belloy et al. 2021).

In humans, QPPs involve the whole brain, including subcortical areas and the cerebellum (Yousefi and Keilholz 2021), and regression of global signal makes QPPs more similar across individuals in terms of the spatial extent of anti-correlations between the DMN and other cortical areas (Yousefi *et al.*, 2018). Global signal regression also reduces variability in the QPPs in mice (Belloy et al. 2018), but it is not yet known whether the pattern extends to subcortical or cerebellar areas. The amplitude of the QPP in subcortical areas is lower than in cortical areas in humans (Yousefi & Keilholz, 2021), and the relatively small amount of data obtained from rats and mice - compared to the large Human Connectome Project database (David C. Van Essen et al., 2013) - may hinder detection. Since surface coils are often used for rodent rs-fMRI studies, there is also a loss of sensitivity with depth that further limits the capability to detect small fluctuations. Finally, many studies in rats and mice do not acquire data from the cerebellum in order to maintain higher spatial and temporal resolution.

The timing of the QPPs varies across species.  In humans, it is approximately 20 seconds in length, while in rodents, the length of the QPP varies depending on the anesthetic used (~5-10 s; Thompson et al. 2014). The variations observed are in line with the differences in lag times between infraslow activity and the BOLD signal, and variations observed in the hemodynamic response to stimulation under similar conditions, suggesting that a similar mechanism accounts for the differences in the length of the QPP (Pan et al. 2013).

*Coactivation patterns.* Coactivation patterns (CAPs) analysis is another method used to capture dynamically evolving brain activity. It was illustrated in (Xiao Liu & Duyn, 2013; Tagliazucchi, Balenzuela, et al., 2012) that, by extracting the time points at which there is high signal intensity in a selected region of interest, spatial patterns similar to resting-state networks seen in functional connectivity-based analysis can be obtained using only a very small fraction of the data. These selected fMRI frames can be further clustered into several groups based on their spatial patterns. Although both sliding window correlation and CAPs use k-means clustering to identify the time-varying features, sliding window correlation applies a clustering algorithm on functional connectivity matrices calculated based on a relatively long time window, which has a much coarser time scale than the spatial patterns in individual frames that are clustered in CAPs. Thus, CAPs are sensitive to faster dynamic activity than sliding window correlation. Using the CAPs method, it was demonstrated that the DMN could be temporally decomposed into multiple DMN CAPs with different spatial configurations (Xiao Liu & Duyn, 2013), suggesting that the DMN and TPN may have a many-to-one correspondence instead of a one-to-one relationship suggested by traditional non-dynamic methods (Xiao Liu, Zhang, et al., 2018). Using the CAPs method in awake and anesthetized rodents, Liang et al. have found that the networks of infralimbic cortex and somatosensory cortex could be decomposed into several spatial patterns in a similar way (Liang et al., 2015). In mice, CAPs occur at specific phases of fMRI global signal fluctuations using rs-fMRI under anesthesia (Gutierrez-Barragan et al., 2019). Zhiwei Ma and Zhang (2018) investigated the temporal transitions among CAPs in both awake rodents and humans and found the transitions are nonrandom in both cases (Zhiwei Ma & Zhang, 2018).  There have also been applications of CAPs in task fMRI, e.g. language tasks and spatial perception tasks (Bordier & Macaluso, 2015) and visual attention tasks (Bray et al., 2015) in humans. We are unaware of similar reports in fMRI studies in rodents.

Like other dynamic approaches that are purely based on fMRI measurement, it is hard to interpret whether the distinct spatial patterns revealed by CAPs are attributed to fluctuations in neural activities or physiological noise, with fMRI being an indirect measurement of neural activities. Applying the CAPs method to animal models, like rodents, can play an important role in investigating the neural correlates of such temporal fluctuations because of the availability of more direct and invasive measurements of neural activity. Zhang et al. have used concurrent local field potential and fMRI measurements to identify potential neural correlates of CAPs on anesthetized rodents, and have found that the CAPs observed in rs-fMRI are linked to the time points with high local field potential broadband power (X. Zhang et al., 2020). To what extent, and by what means the temporal fluctuations are linked to neural activity remains to be discovered. Figure 4 shows examples of *sliding window correlation, quasi-periodic patterns, and coactivation patterns.* Please see a full description of the technicalities and parameters used for these three methods in Fig. S2—S5 in the Supplementary Materials.

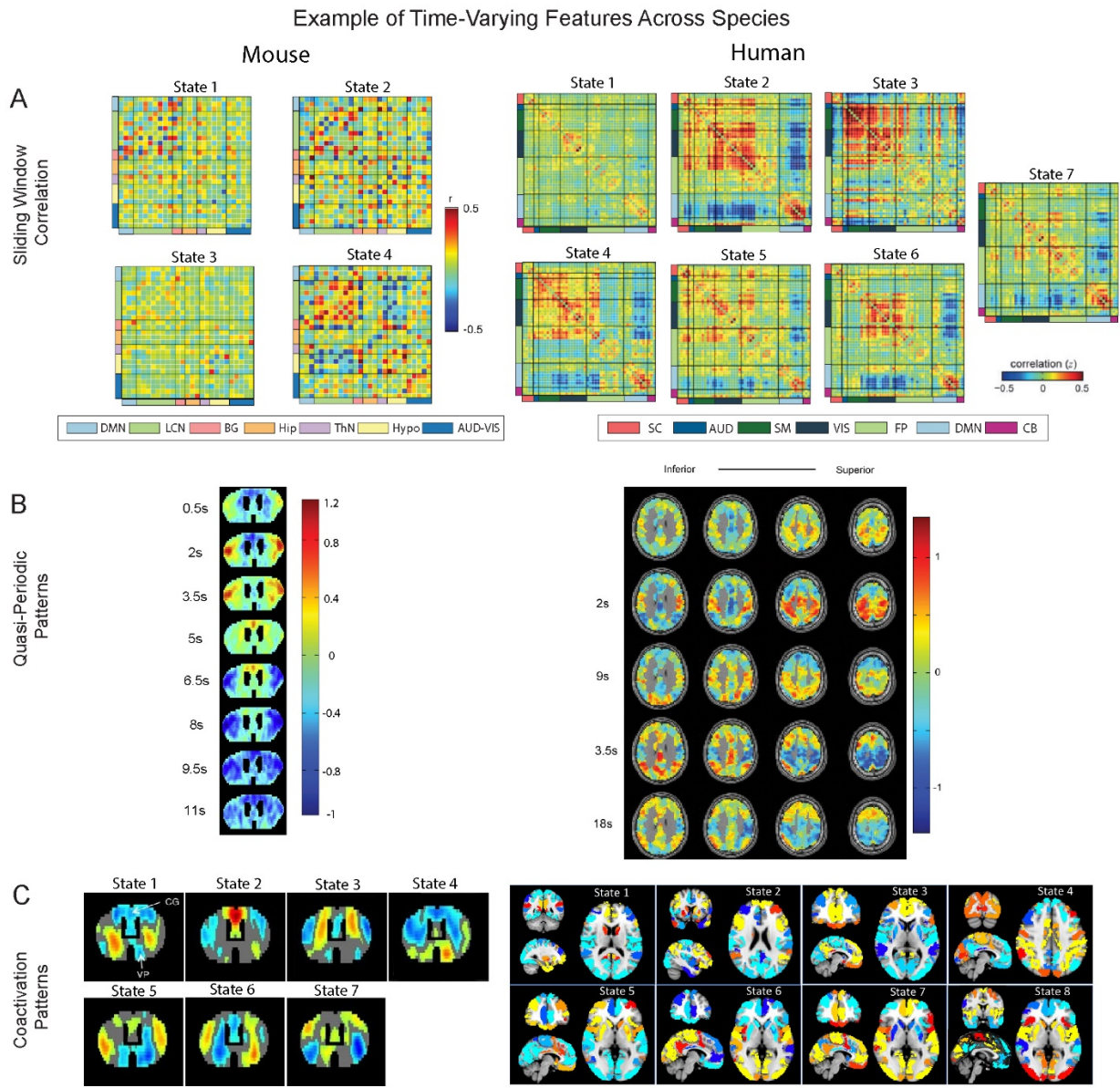

**Figure 4:** *Overview comparison of windowed approaches, quasi-periodic patterns, and coactivations patterns. (A) Distinct brain states revealed by sliding window correlation are from* (Tsurugizawa & Yoshimaru, 2021) *for mice (left) and from* (Allen et al., 2012) *for human brains (right). Mice networks (left) include DMN (light blue), lateral cortical network (LCN, light green), audio-visual network (AUD-VIS, dark blue), subcortical basal ganglion (BG, coral), hippocampus (Hip, orange), thalamus (ThN, light purple), and hypothalamic network (Hypo, light yellow). For humans (right), a similar set of color codes indexing networks that are homologous to the mice counterparts are also added to the figure. Specifically, human networks (right) include DMN (light*

*blue), frontoparietal (FP, light green), somatomotor (SM, dark green), auditory (AUD, blue), visual (VIS, dark blue), subcortical (SC, light red), and cerebellar (CB, violet). Note that the central executive network, which is known as FP in humans, is referred to as LCN network in mice* (Gozzi & Schwarz, 2015). *(B) The quasi-periodic patterns and their correlation with the whole brain images across time are from (Belloy et al. 2018) for mice (left) and from* (Majeed et al., 2011) *for humans (right). (C) Brain states determined by coactivation patterns are from* (Adhikari, Belloy, Van der Linden, Keliris, Verhoye, et al., 2021) *for wild-type mice (left) and from* (Janes et al., 2020) *for human brains (right). CG and VP denote the cingulate cortex and ventral pallidum, respectively.*

## VI. Applications

*Studies of cognitive processes.* Cognitive neuroscience has made extensive use of fMRI to study the mechanisms underlying cognitive processes like memory, learning, and attention. In rodents, the main barrier to similar studies is the common use of anesthesia, which makes it impossible to perform similar behavioral studies in the scanner. However, using rs-fMRI, researchers are able to examine functional connectivity before and after a variety of behavioral training.  In this case, changes that occur during the training itself are not observed and anesthesia can be used during the scans. In an *in vivo* study from 2016, Nasrallah et al. reported that functional connectivity changes can be observed between the hippocampal CA3 and thalamus, septum, and cingulate cortex in sedated rats that had undergone a cognitive task including training on a hidden platform water maze (Nasrallah et al. 2016). Another possibility is to use unanesthetized animals.  For example, Liang et al. observed the effects of predator odor exposure in an inescapable environment on freezing behavior, anxiety levels, and rs-fMRI seven days following the occurrence of the event. They determined that a single traumatic episode can be detected in the "compromised" network of the amygdala and prefrontal cortex even a week following the event (Liang et al., 2014).

*Relationship to neural activity.* The use of rodent models allows the direct investigation of the relationship between BOLD and neural activity using simultaneous noninvasive (e.g., fMRI or hemoglobin imaging) and invasive (e.g, local field potentials or calcium imaging) recordings (Drew et al., 2020; Shella D. Keilholz, 2014; Logothetis et al., 2001; Magri et al., 2012; Mitra et al., 2018; W. ju Pan et al., 2011; Schwalm et al., 2017; Shmuel & Leopold, 2008; Garth J Thompson et al., 2012; Garth John Thompson, Merritt, et al., 2013). These studies have helped to elucidate the coordinated neural and vascular oscillations that give rise to functional connectivity (Drew et al., 2020; Logothetis et al., 2001; Magri et al., 2012; Schwalm et al., 2017; Shmuel & Leopold, 2008). Studies have shown that different BOLD dynamic patterns tie to neural activity at different frequency bands (S D Keilholz et al., 2013; Shella D. Keilholz, 2014; Pais-Roldán et al., 2021; Garth John Thompson et al., 2015). The infraslow activity (<0.1Hz), in particular, has a distinctive spatiotemporal signature, which does not appear in higher frequency activities (Mitra et al., 2018; W. J. Pan et al., 2013; Garth J Thompson et al., 2012), whereas the higher frequency bands are tied to time-varying functional connectivity (Garth John Thompson et al., 2015).

In humans, the relationship between the BOLD signal and neural activity was studied using either separate or simultaneous recordings of EEG/ECoG/MEG and fMRI (E. Allen et al., 2018; C. Chang et al., 2013; Grooms et al., 2012; Tagliazucchi, Von Wegner, et al., 2012). For example, Kucyi and his colleagues studied the correspondence between the temporal profiles of ECoG frequencies and functional brain networks detected by fMRI (Kucyi et al., 2018). Their results indicated that functional connectivity reveals divergence or unique temporal profiles at different frequency bands, which agrees with the previous discovery in rat somatosensory cortex using simultaneous BOLD functional connectivity and intracranial recordings (Garth John Thompson, Merritt, et al., 2013). Other studies discovered that ECoG functional connectivity at different frequency bands corresponds to different functional structures detected by BOLD (Hacker et al., 2017; Hipp & Siegel, 2015). This relationship in humans was further supported by studies using simultaneously-recorded EEG and fMRI, which have found associations between BOLD functional connectivity and changes in power across multiple frequency bands including delta, theta, alpha, beta, and low-gamma (E. Allen et al. 2018; Tagliazucchi, Von Wegner, et al. 2012; Chang et al. 2013). However, due to the nature of the poor spatial resolution of EEG (roughly 6-20 $cm^2$ of the synchronized cortical area is needed for scalp EEG spike detection) (Singh, 2014), these studies are unable to directly link the electrophysiological basis of spatially specific variations in coupled activity between brain regions for functional connectivity. Studies using MEG have observed time-varying interregional correlations of band-limited power that have similar spatial topography to BOLD functional connectivity networks (Pasquale et al., 2010; Vidaurre et al., 2018).

*Animal models of brain disorders.* The use of animal models to study human neurological disorders is advantageous for many reasons, some of the most notable being the many possibilities afforded by genetic manipulation (McGraw et al., 2017), the greater control of subject variables, and the experimental flexibility that comes from investigative techniques unable to be used in humans, such as the combination of chemogenetics and MRI (Valerio Zerbi et al., 2019). Rodent models in particular have shown great promise in recent years due to their utility in studies involving genetic alteration and their various similarities to the human brain. More specifically, rodent models for diseases characterized by a variety of genetic abnormalities, in which the model isolates a specific genetic manipulation. The utilization of such models allows for the study of phenotypical changes in functional connectivity as the result of a specific genetic modification. A great example of this is seen in studies of Autism Spectral Disorder (ASD) where genetic alterations play a large role in the development of the disease (Sandin et al., 2017). In such efforts, the use of ASD rodent models containing gene-specific mutations has proven incredibly useful in the study of underlying dysfunctions in brain connectomes and functional connectivity (Pagani et al., 2021; V. Zerbi et al., 2021). However, choosing a rodent model must be done carefully as many variables, such as the difference in brain structure and function between rodents and humans as well as the type of rodent used (Ellenbroek & Youn, 2016; Jonckers et al., 2011), act as contributing factors to the efficacy of the model. Rodents have served as the foundation for models of a vast number of diseases, many of which have shown alterations in the DMN in humans (The Significance of the Default Mode Network (DMN) in Neurological and Neuropsychiatric Disorders: A Review, 2016). The finding that rodents exhibit a homologous DMN network consisting of both cortical and subcortical structures supports their use as a model for brain disorders. There already exist various rodent disease models for disorders such as Alzheimer's Disease (Braidy et al., 2011; Puzzo et al., 2015), Parkinson's Disease (Campos et al., 2013; Harvey et al., 2008; Terzioglu & Galter, 2008), Epilepsy (Curia et al., 2008; Sharma et al., 2007), Attention-deficit/hyperactivity disorder (ADHD) (Sagvolden et al., 2005), Schizophrenia (Powell & Miyakawa, 2006; Pratt et al., 2012), Post-traumatic stress disorder (PTSD) (Flandreau & Toth, 2018; Verbitsky et al., 2020), as well as Depression and other mood disorders (Gururajan et al., 2019; Pollak et al., 2010). However, in this paper, we will focus mostly on those used for Alzheimer's disease, Epilepsy, ADHD, and Autism spectrum disorder (ASD).

*Alzheimer's disease.* Alzheimer's disease (AD) is a neurodegenerative disease that accounts for the majority of dementia cases (Guerreiro et al., 2020). A fully comprehensive and translatable AD rodent model does not exist due to many factors, some of the most prominent being that AD is a multifaceted, complex disease that is highly variable and heterogeneous across patients; also, the fact that there exist different types of AD and replicating their progression in rodents is difficult (Drummond & Wisniewski, 2017). AD is characterized by the accumulation of ß-amyloid (Aß) into Aß plaques and the presence of tau neurofibrillary tangles throughout the brain. These pathologic hallmarks are then followed by clinical symptoms including cognitive decline and memory loss ("2020 Alzheimer's Disease Facts and Figures," 2020). The interaction and complex progression of these two pathologic elements are not fully understood, although they are the basis of many AD models used to date.

Many studies in humans have shown alterations in functional connectivity and network dysfunction in brains that have been affected by AD. Abnormal DMN connectivity, especially decreased DMN connectivity, has been shown to occur in prodromal and later stages of AD (Adriaanse et al., 2014; Badhwar et al., 2017; Gour et al., 2014; Jacobs et al., 2013), with reduced DMN connectivity being thought to correlate with AD severity (Brier et al., 2012; Petrella et al., 2011; Zhou et al., 2010). Hyperconnectivity in both the Salience and Limbic networks has also been shown in humans with AD (Badhwar et al., 2017; Gour et al., 2014; Peterson & Li, 2018). Similarly, decreased DMN connectivity (Tudela et al., 2019) and general functional connectivity abnormalities (Anckaerts et al., 2019) have been shown in TgF344-AD rats. Interhemispheric hippocampal functional connectivity has been shown to decrease in 3xTg-AD transgenic mice when compared to controls (Manno et al., 2019) and decreased functional connectivity was also shown to correlate with tau protein expression in transgenic mouse models of tauopathy (Green et al., 2019). These findings are mostly congruent with alterations reported in humans.

*Epilepsy.* Epilepsy is a brain disorder characterized by the excessive firing of neurons causing recurrent seizures and is a condition of altered neural network organization (Dumlu et al., 2020; Kramer & Cash, 2012; The Significance of the Default Mode Network (DMN) in Neurological and Neuropsychiatric Disorders: A Review, 2016). A recently published systematic review of 24 studies looking at RSN changes in idiopathic generalized epilepsy (IGE) reported reduced functional connectivity in DMN in IGE patients compared to healthy controls, regardless of the subtype of IGE. Several studies also found changes in non-DMN local connectivity, including hyperconnectivity in the cerebellum and hypoconnectivity in the left middle and inferior temporal gyri involved in the dorsal attention network, the precentral and postcentral gyri of the sensorimotor network (Parsons et al., 2020). However, in rat models of absence epilepsy (WAG/Rij), increased interhemispheric connectivity in the somatosensory cortices was found (Mishra et al., 2011). Similarly, Pedersen et al. found increased connectivity within and between somatosensory cortices, putamen, and thalamus in patients with the *GABRG2* genetic variant, which is known to cause early-onset absence epilepsy and febrile seizures (Pedersen et al., 2019). The variable findings point to the importance of understanding the diversity in etiologies that lead to different pathological processes in epilepsy.

To focus on a relatively less heterogeneous group, we turn to studies of temporal lobe epilepsy (TLE). It is the subtype of epilepsy that has been studied most extensively in relation to disruptions in resting-state brain connectivity, the DMN in particular (The Significance of the Default Mode Network (DMN) in Neurological and Neuropsychiatric Disorders: A Review, 2016). As in IGE patients, TLE patients have been shown to have decreased connectivity in the DMN (Haneef et al., 2012, 2014; Liao et al., 2010; Pittau et al., 2012; Roger et al., 2020). In addition to global network connectivity, also of interest in focal epilepsy is the regional connectivity in epileptogenic or seizure-onset zones. Studies have found conflicting results. With regions of interest defined including the bilateral hippocampi, Haneef et al. found increased connectivity in TLE patients in regions of the limbic system, including bilateral temporal lobes, insula, fornix, angular gyrus, and thalamus (Haneef et al., 2014). A recent study by Roger et al., which focused on examining the language and memory networks in mesial TLE patients, also found increased connectivity in limbic regions around the dysfunctional hippocampus (Roger et al., 2020). In contrast, Pittau et al. and Bettus et al. found decreased connectivity in the epileptogenic areas, including the hippocampi and amygdala (Bettus et al., 2009; Pittau et al., 2012). In a study of lithium-pilocarpine (LIP)-treated rat models of TLE, Jiang et al. showed a decreased connectivity in the hippocampal network including the hippocampus, thalamus, retrosplenial cortex, and somatosensory cortex (Jiang et al., 2018). On the other hand, increased connectivity was seen between the hippocampal network and the visual cortex, mesencephalon, and insula. Functional connectivity within the hippocampal network was positively correlated with spatial memory task performance and the connectivity between hippocampal network and visual cortex was negatively correlated with object memory performance. Christiaen et al. (2019) used an intraperitoneal kainic acid (IPKA) rat model of TLE to longitudinally study functional connectivity changes during the development of epilepsy (Christiaen et al., 2019). The group found decreasing connectivity most evident in the retrosplenial cortex, which is a region of the DMN in rats, over time following one episode of induction of status epilepticus.

Taken together, studies have shown the utility of rs-fMRI in understanding the disruptions in networks in epilepsy and possible correlations with neurocognitive functioning. Varying results among studies is a reflection of the heterogeneity of the disorder, encompassing diverse etiologies and pathologies under the same diagnosis. Given the variability in connectivity changes both at the group and individual levels, longitudinal studies tracking change over time as well as those comparing pre- and post- treatment would be crucial to support using functional connectivity measures as a biomarker in epilepsy.

*Attention-Deficit/Hyperactivity Disorder (ADHD).* Studies of brain functional connectivity in ADHD have commonly involved the DMN, frontoparietal, and corticostriatal networks. A systematic review by Posner et al. revealed that in ADHD, the negative correlational relationship between DMN and frontoparietal is either diminished or absent, and may represent the neural substrate for an interruption in attentional control (Posner et al., 2014). They also reviewed studies that reported hypoconnectivity within the DMN.  In rodents, Huang et al. compared DMNs in the spontaneously hypertensive rat (SHR) strain, a commonly used rodent model for the combined subtype of ADHD, and a genetically related Wistar Kyoto rat (WKY) control strain (S. M. Huang et al., 2016). Compared to controls, SHR rats displayed hyperconnectivity in the DMN overall, which is not in line with results from many human studies. The majority of studies included in the aforementioned systematic review and meta-analysis were in children. The conflicting findings of connectivity changes point to the importance of understanding the heterogeneity in the patient population as well as different rodent models of ADHD. For example, results from SHR rats may align with those seen in adults with the combined type of ADHD, but not with findings in children and/or different subtypes. Future studies in rodent models of ADHD exploring these sources of sample heterogeneity—subtypes, age group, and medication effects—would be valuable in better understanding the neuropathology of ADHD and mechanisms of treatment, as well as discovering new potential targets of treatment.

*Autism Spectrum Disorder (ASD).* A systematic review of rs-fMRI studies in ASD revealed both hypoconnectivity and hyperconnectivity in and between different regions and networks (Hull et al., 2016). A finding commonly reported by numerous studies was global hypoconnectivity in the DMN. However, studies found both decreased and increased connectivity in different regions of the DMN that also varied with the age of the participants (Lucina Q Uddin et al., 2010; Washington et al., 2014). A recent meta-analysis by Lau et al. of rs-fMRI studies of ASD that used non-seed-based methods revealed that individuals with ASD exhibit a reduced local functional connectivity in the dorsal posterior cingulate cortex (dPCC) as well as an area that extends from the postcentral gyrus in the right medial paracentral lobule (Lau et al., 2019). The authors did not identify any regions of hyperconnectivity in the ASD group compared to controls. The same group also conducted a meta-analysis of seed-based studies of connectivity involving the PCC in ASD and showed that there is hypoconnectivity between the PCC and ventromedial prefrontal cortex in individuals with ASD compared to healthy controls (Lau et al., 2020).

Various mouse models of genetic mutations known to cause autism have been studied and were also found to exhibit hypoconnectivity. A mouse model of ASD with a homozygous deletion of *Cntnap2* was found to have decreased connectivity between the cingulate cortex and the retrosplenial cortex, which represent the anterior component of the mouse DMN and a posterior hub, respectively (Liska et al., 2018). The DMN hypoconnectivity was correlated with reduced social behaviors in mice, which is in line with human studies that found a negative correlation between DMN connectivity and social and communication impairments in individuals with ASD (Assaf *et al.*, 2010; Bertero *et al.*, 2018; Liska *et al.*, 2018; Liu and Huang, 2020).

Bertero et al. found reduced global connectivity in the medial prefrontal cortex in both human carriers and mice with 16p11.2 deletion, one of the most common copy number variations (CNVs) in ASD (Bertero et al., 2018). This mouse model was also found to have significantly reduced anterior-posterior connectivity along the axis of the cingulate and retrosplenial cortex. In 2020, Tsurugizawa et al. developed and used a system to obtain awake fMRI in mice with 15q duplication, a model for another well-known CNV in ASD (15q11-13 duplication) (Tsurugizawa et al., 2020). When compared to wild-type mice, *15q dup* mice had widespread resting-state hypoconnectivity involving all 32 selected regions of interest playing a role in cognitive function and odor recognition.

In accordance with earlier findings of variable connectivity alterations in ASD, a recent bi-center study of functional connectivity in 16 different mouse models of ASD, including mice with deletions of 16p11.2, *Fmr1*, *Cntnap2*, and *Cdkl5*, found a wide spectrum of altered connectivities (V. Zerbi et al., 2021). The authors emphasize that the lack of a central pattern of aberrant connectivity in ASD highlights the vast heterogeneity of the disorder. Interestingly, however, different genetic models of ASD etiology could be grouped into four distinct clusters based on network connectivity alterations. The existence of distinct subtypes of rs-fMRI connectivity patterns in ASD points to the need for individualized approaches to treatment (V. Zerbi et al., 2021).

Studies examining rodent models and humans with the same genetic mutations displaying comparable phenotypes of ASD are valuable in understanding the neuropathology of the disorder as well as potential targets for treatment. However, it is also crucial to acknowledge the diversity in the etiologies of ASD since only about 10% of ASD cases can be attributed to monogenic mutations (Liska & Gozzi, 2016). As in the name, individuals with ASD display a wide spectrum of levels of functioning, with varying severity of symptoms. Functional connectivity alterations may serve as unique neural biomarkers in different subgroups of ASD.

## VII. Open questions and challenges

*Protocol Differences Produce Challenges of Reproducibility*. Over the last decade, large, publicly available datasets have become the expected procedure for human resting-state fMRI studies (B. B. Biswal et al., 2010; Milham et al., 2018). The efforts towards centralized and publicly-available databases have accelerated the pace of research on the systems-level activity of the brain. Many repositories include data acquired at different sites with different protocols while others obtain data from many subjects using the same protocol at one or a few sites: BOLD5000 (N. Chang et al., 2019), NKI-Rockland sample (Nooner et al., 2012), Adolescent Brain Cognitive Development study (ABIDE I & II) (Casey et al., 2018; Di Martino et al., 2014), the UK Biobank (Littlejohns et al., 2020), Open Access Series of Imaging Studies (OASIS) Cross-Sectional (Marcus et al., 2007), OASIS Longitudinal (Marcus et al., 2010), OASIS-3 (LaMontagne et al., 2019), and the Human Connectome Project (David C. Van Essen et al., 2013). Efforts by Markiewicz et al. (2021) have accelerated the centralization of smaller datasets into one database where over 17,000 participants across 493 datasets consisting of a mix of MRI, EEG, MEG, and positron emission tomography (PET) data are publicly available to use through the use of a common community standard increasing the utility of shared data to study brain functional connectivity across a wide array of data and datatypes (Markiewicz et al., 2021). For rodent rs-fMRI, the situation is akin to the situations in humans a decade ago, with a few shared datasets, many of which were acquired under different conditions, and none of which contain the sheer numbers of subjects that are available in human databases (S.-H. Lee et al., 2021; Y. Liu et al., 2020). The field is ripe for the type of large-scale studies that are performed in humans.

One of the benefits of large-scale studies like the Human Connectome Project is the widespread adoption of specific imaging protocols. In rodents, there is still no standard protocol, and acquisition parameters such as spatial resolution, sampling rate, and coverage vary substantially from lab to lab. Moreover, the situation is complicated by the use (or lack) of anesthesia. Both the type of anesthetic and the delivery protocol also vary across studies. The move to unanesthetized animals would mitigate this aspect of site-to-site variability, but imaging awake rodents remain challenging and less robust (Grandjean et al., 2020).

In humans, most studies use one of a handful of atlases to parcellate the brain. Standard atlases remain to be adopted in rodents, although some excellent candidates are emerging (Barrière et al., 2019; Grandjean et al., 2020; Zhiwei Ma & Zhang, 2018). The ability to obtain high-quality images of the majority of the brain is a fairly recent development in rodents, which partially explains the lag relative to human studies.

*Variability within and across animals.* Although neuroimaging studies from rats and mice are often lumped together in the single category of "rodents", there are substantial differences across the species that have barely been explored. Even more, there are differences between strains for each species (Walkin et al., 2013; Yoon et al., 1999). While strain differences are sometimes exploited to examine the genetic basis for functional differences, they are most often ignored. Finally, even within a strain, many studies use a single sex to minimize unwanted differences (in susceptibility to anesthesia, for example). In humans, it is standard practice to track sex as a potential covariate of interest, and future work in animals should do the same.

*The limitations of animal models.* It is well known that animal models often do not capture key aspects of human cognition and disorders such as neurodevelopmental disorders (Zhao & Bhattacharyya, 2018); multiple sclerosis (Denic, Johnson, et al., 2011); neurological and psychiatric disorders (Fernando & Robbins, 2011; Monteggia et al., 2018); psychoactive drug use and addiction (Müller, 2018; Venniro et al., 2020); and obesity and diabetes modeling (Kleinert et al., 2018) where rodents do not exhibit similar human behavior coinciding with the disease. Indeed, the prior section describes cases where findings in humans and rodents are similar and other cases where the findings are discordant. In cases where results differ substantially in humans and rodents, the underlying cause may be an inadequate rodent model of the disorder, a failure to capture the heterogeneity present in the population, or a disconnect between the mechanisms in rodents and humans. Similarly, while there are obvious parallels between rodents and human brains, there are also major differences that challenge neuroscience research. For sensory areas, there is a relatively simple way to determine that function is similar across species, but for higher-level areas that integrate input and contribute to high-level cognition, comparisons are less straightforward. For example, the default mode network is implicated in a wide variety of human behavior, and an analogous network has been proposed for rodents. How, though, do we determine if the network function is truly comparable? This is an ongoing challenge for animal research that is intended to inform or guide the interpretation of work in humans.

*Open Questions.*

*What causes the widespread coordination that we observe as functional connectivity?* This is the fundamental question for rs-fMRI researchers, and animal studies play a key role in elucidating the underlying mechanisms using multimodal, invasive measurement techniques as well as non-invasive fMRI. At the most basic level, it appears that the general mechanisms that relate neural activity to the changes in blood oxygenation measured with fMRI are similar across species, which increases confidence that work in animals can help to interpret work in humans. Recent advances in the evolutionary study have enabled quantifying the functional homology across animals and humans through a function-based method for cross-species alignment (Xu et al., 2020). Focusing on the evolutionarily conserved networks, the salience, DMN, and central executive networks (which is frontoparietal in humans), researchers have been able to convert mouse brain maps into human standard coordinates and open the door to fully data-driven comparisons across rodents and humans by leveraging optogenetics in mice and tractography in humans (Mandino et al., 2021). However, much is still unknown, and it is easy to imagine that the frequencies of neural activity, the precise mechanisms of neurovascular coupling, and the brain states of neuromodulatory input mechanism might have important differences across species. We can only continue to investigate these relationships and the conditions under which they hold, in an attempt to identify commonalities that might be tested in a subset of human studies.

*How do we obtain results that are most comparable across species?* Certainly, robust protocols for imaging unanesthetized animals would be a large step towards improving comparability. In the meantime, obtaining data under multiple contrasting anesthetic conditions (Flatt et al., 2021; Mathew et al., 2021; Stenroos et al., 2021; Williams et al., 2010) can help to identify features that are anesthesia-specific and less likely to be conserved in awake animals, and common features that are more likely to be conserved. The use of awake animals also opens the door for some specific behavior-related brain activity studies, in addition to a common sensory stimulation study paradigm.

Another important issue is that motion artifacts must be minimized in both species. The human fMRI studies are mostly relying on participants' cooperation to avoid voluntary head motion along with immobilizing assistance. Rodent fMRI are mostly relying on anesthesia and ventilation, but head motion spikes were often observed in the awake state due to the animal struggling in the fixed position, which is even though reduced a lot over the training session. Notably, the subjects' motion may be relevant to stress levels and brain physiological and functional state variations that would add complicated situations in functional brain activity. These unwanted physiological contributions can affect the images in each species in different ways, so careful cleanup is needed (Drew et al., 2020). With these many limitations, these open questions are paramount for understanding the use of rodent modeling precluded in studies in human functional connectivity.

*Conclusions.*

Rodent models have played and will continue to play an important role in understanding the functional architecture of the brain as measured with rs-fMRI. While differences in acquisition strategies, particularly the use of anesthesia, make comparisons between rodents and humans less than straight-forward, the observation of comparable networks across species increases confidence that multimodal studies in animal models can help to interpret rs-fMRI in humans. Research in rodent models has already established a firm neural basis for both time-averaged and time-varying characterizations of functional networks, and further investigation will provide more insight into the mechanisms that coordinate systems-level neural activity and hemodynamics. Resting-state fMRI is already being applied in animal models of human disorders and, in some but not all cases, has detected alterations consistent with those observed in the patient population. Further efforts to understand the correspondence of rs-fMRI across species and the development of improved animal models may eventually allow researchers to probe the mechanisms that underlie alterations in functional connectivity.

**Acknowledgments**

The authors would like to thank Harry Watters for helpful discussions, and acknowledge funding from 1R01MH111416, 1R01NS078095, and 1R01EB029857. TJL would like to thank the financial support from the George R. Riley Fellowship.

**Supplementary Materials**

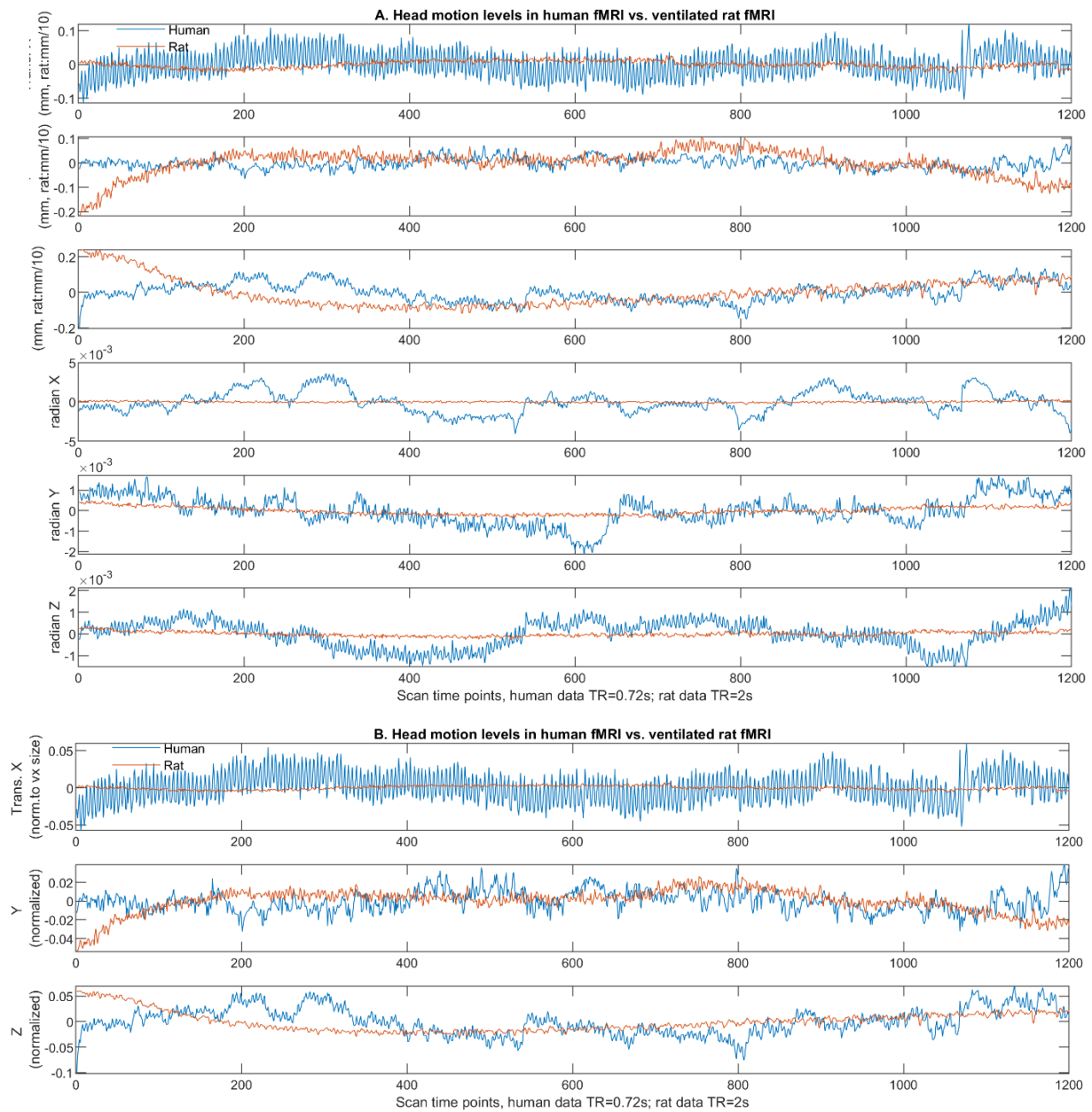

**Figure S1.** Head motion levels in a representative human rs-fMRI data set v.s. rat rs-fMRI data set. The blue lines show translation and rotation for a randomly selected subject from the 3T Human Connectome Project dataset (Van Essen et al. 2013); voxel size is 2mm isotropic. The red lines show translation and rotation for a randomly selected rat imaged under dexmedetomidine and low-dose isoflurane while paralyzed and mechanically ventilated; voxel size is 0.4mm isotropic (x10 scale in shown). The representative rat data are from (Pan et al. 2020). The full set of rat data will be deposited along with the project (https://github.com/grandjeanlab/MultiRat). The linear drifts were removed. Relative to human rs-fMRI data, the typical rodent rs-fMRI data has an extremely low level of head motion for both translation and rotation. Breath-related motions are common in human rs-fMRI but rarely seen in the data from ventilated rats. (A) shows the motion comparison in absolute translation distance and rotation (human, 2mm isotropic voxels; rat, 0.5mm isotropic voxels). The motion level in X direction translation (mean in mm +/- std), human: 0.0333+/-0.0227, Y, 0.0180+/-0.0140, Z, 0.0425+/-0.0004; rat: X 0.0009+/-0.0006, Y, 0.0038+/-0.0036, Z, 0.0062+/-0.0000; The motion level in X-direction rotation (mean in radian +/- std), human: 0.0011+/-0.0009, Y, 0.54e-3+/-0.41e-3, Z, 0.52e-3+/-0.36e-3; rat: X 0.0001+/-0.0001, Y, 0.16e-3+/-0.10e-3, Z, 0.09e-3+/-0.06e-3. Relative to the ventilated rat data, the human data may have significantly higher levels of rotation in all 3 directions, especially one direction with possibly breath waves. (B) shows the motion comparison in voxel relative sizes (normalized to absolute sized in each direction correspondingly). The motion level in X translation (mean+/-std), human: 0.0167+/-0.0114, Y, 0.0090+/-0.0070, Z, 0.0213+/-0.0155; rat: X, 0.0022+/-0.0014, Y, 0.0095+/-0.0089, Z, 0.0154+/-0.0116; The motion level in X direction (probably respiration-related) in human data may be almost one order higher than ventilated rat data. In the other two directions, the translation levels are similar or slightly better in the ventilated rat after linear drift removal (if higher-order drifts are removed, the rat head translations would be much smaller than the human translations).

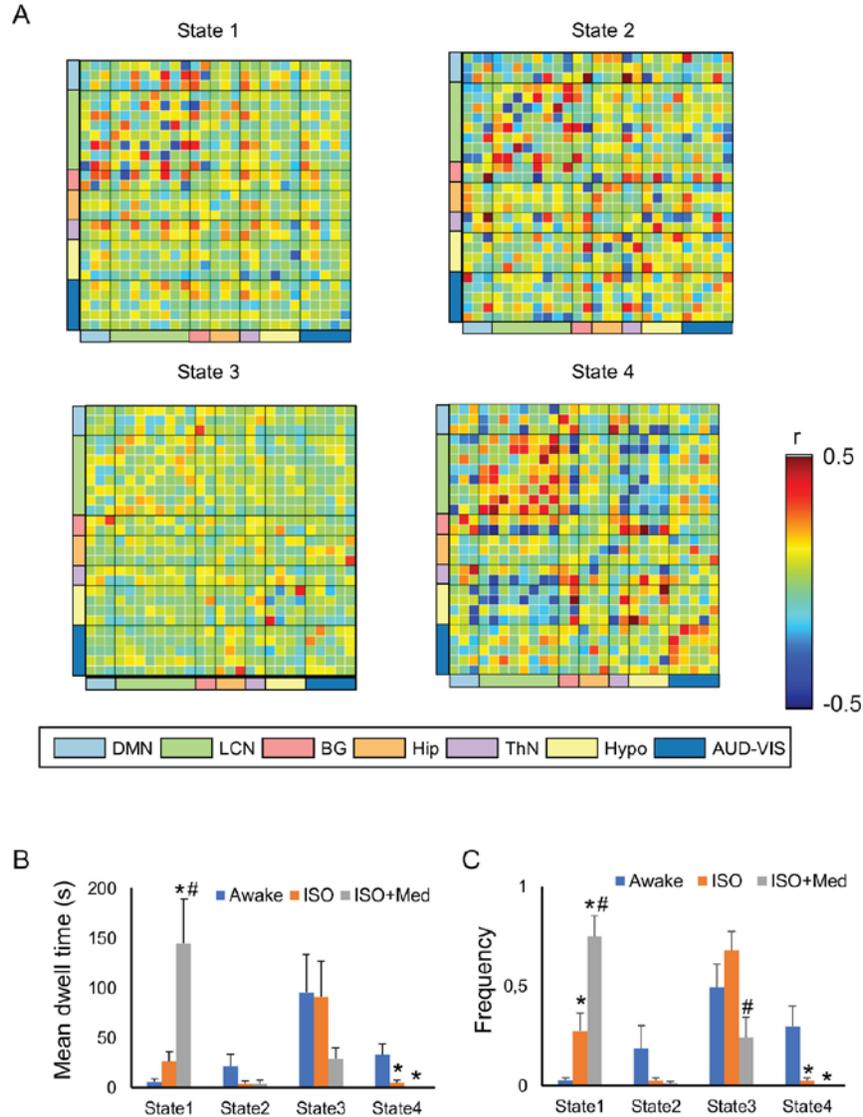

**Figure S2**: Example of windowed approaches for mice. Distinct brain states revealed by sliding window correlation are excerpted from (Tsurugizawa and Yoshimaru 2021) for mice. Four distinct mice brain states (A) were determined through the following procedures: First, using the fMRI ICA Toolbox (GIFT v3.0b http://trendscentre.org/software/gift/), fMRI timeseries for all three mice groups (i.e., 1 awake group (n=9), 1 isoflurane anesthesia group (n=9), and 1 medetomidine +isoflurane anesthesia group (n=9)) were extracted from ROIs that are determined by the ICA. These ICA determined ROIs cover the following brain networks: auditory-visual network (AUD-VIS); subcortical basal ganglion network (BG); default mode network (DMN); hippocampus (Hip); hypothalamus (Hypo); lateral cortical network (LCN); thalamic network (ThN). Second, dynamic FC matrices were computed among these extracted fMRI timeseries using GIFT Dynamic FNC Toolbox (v1.0a). The window length was set to be 45 s (30 TRs), which slid in steps of 1.5 s (1 TR). Third, GIFT k-mean clustering was performed on these dynamic FC matrices to determine the optimal brain states. In this study, four optimal distinct brain states were found (K=4). (B and C) Averaged dwell time (B) and frequency of occurrence (C) of each brain state are also measured and compared between different groups using GIFT Dynamic FNC Toolbox (v1.0a). The dwell time measures the frequency of a state unchanged between the current and the next window, whereas the frequency of occurrence measures the number of windows in each state. Here, *p < 0.05 compared with the awake condition in each state, Tukey-Kramer test (df = 17). #p < 0.05 compared with the Iso condition in each state, Tukey-Kramer test (df = 17).

# Sliding Window Correlation Example on Humans from Literature

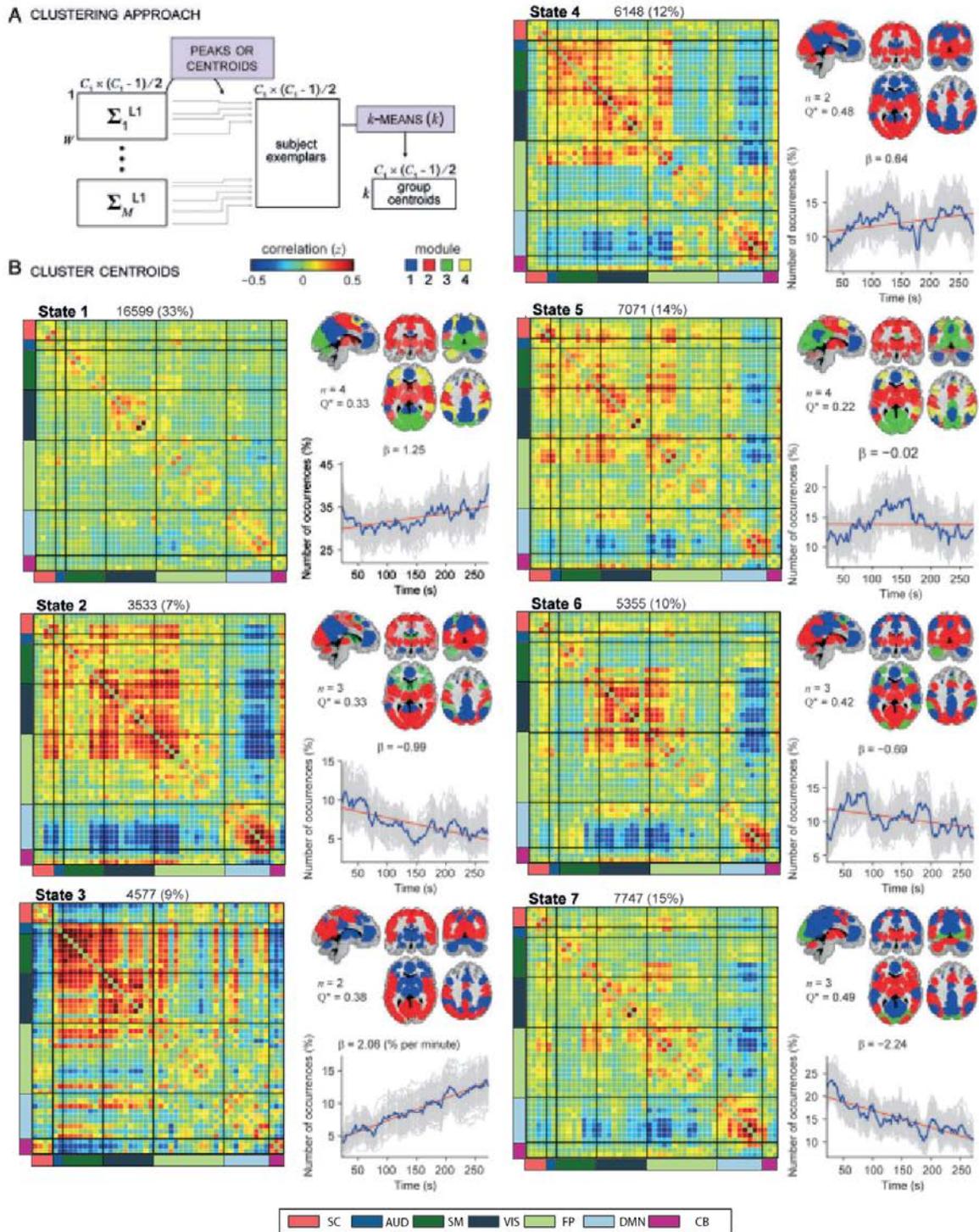

**Figure S3.** Example of windowed approaches for humans. Distinct brain states revealed by sliding window correlation are from (Fig. 5, Allen et al. 2012) for human brains. Seven distinct human brain states (B) were determined from the calculated sliding-window correlation matrices through the K-mean clustering procedure (A). More specifically, a group-level spatial ICA was first

performed to determine the ROIs for all group data (405 healthy participants, 200 females, with age range 12–35 yro, and mean age at 21.0 yro). Second, dynamic FC matrices were computed among these extracted fMRI timeseries. The window length was set to be 44 s (22 TRs), which slid in steps of 2 s (1 TR). Third, GIFT K-mean clustering with 500 repetitions was performed on these dynamic FC matrices (as shown in (A)) to determine the optimal clusters and the cluster number. Each brain state is the K-mean cluster that was summarized by the cluster centroid (see the FC matrix as shown in (B)). In this human study, seven optimal distinct brain states were found (K=7). The number of occurrences of each state is shown as a function of time at the bottom right of each FC matrix. Finally, the Louvain algorithm was repeated on 100 bootstrap resamples within each cluster to obtain the modular structure for each brain state (top right). The module colors (red, blue, green, and yellow in the brain plots) were matched across states such that similar partitions share the same color. In order to compare with the mice brain networks in Figure S2, we added a similar set of color codes indexing networks that are homologous to the mice counterparts, which include DMN (light blue), frontoparietal (FP, light green), somatomotor (SM, dark green), auditory (AUD, blue), visual (VIS, dark blue), subcortical (SC, light red), and cerebellar (CB, violet). Note that the central executive network, which is known as FP in humans, is referred to as the LCN network in mice (Gozzi and Schwarz 2015).

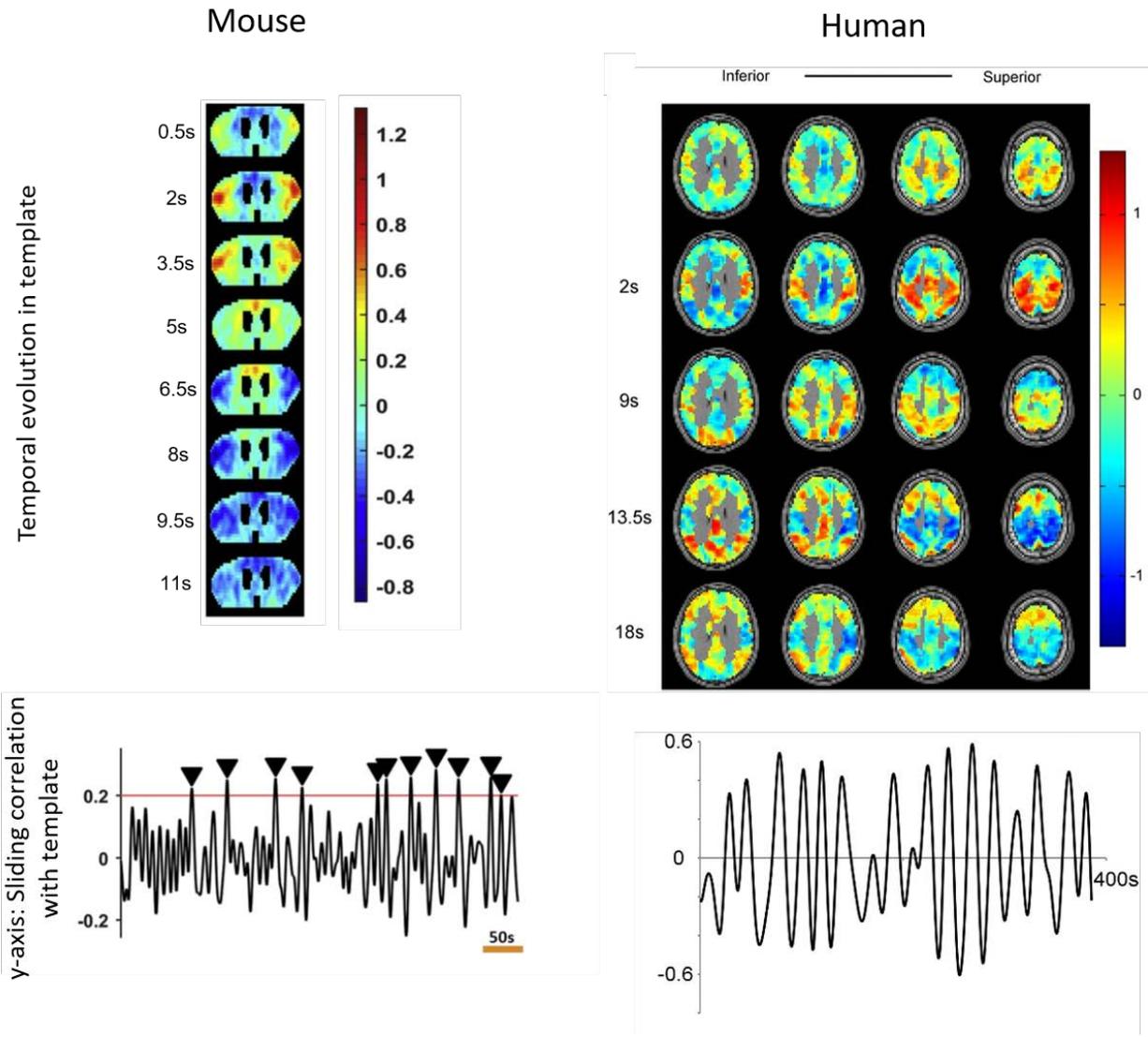

**Figure S4.** Example comparison of quasi-periodic patterns in mice and humans. The quasi-periodic patterns (QPPs; upper) and their correlation with the whole brain images across time (bottom) are from (Belloy et al. 2018) for mice (left) and from (Majeed et al. 2011) for humans (right). Specifically, the upper row illustrates the temporal evolution of the detected QPPs at an interval of 1.5s for mice (left) and an averaged interval of 5.67s for humans (right). For both mice and humans, alterations between DMN and task-positive network (TPN) are seen. For example, at the beginning stage of QPPs (t=2s), the TPN for both mice (the lateral cortical network) and humans (the attention network) shows a high signal when the signal from DMN is reduced. At an intermediate stage (t=5s for mice and t=9s for humans), a reduced TPN and an increased DMN are shown for both mice and humans. At the final stage (t=6.5s for mice and t=13.5s for humans) positive intensity in DMN and negative intensity in the TPN are observed. The signal intensity returns to baseline after some time, as seen in the last frame (t=11s for mouse and t=18 s for human). Separately, the bottom row illustrates the correlation value between the detected QPP and the whole brain images over time, namely 'sliding correlation with template' in the figure. The peaks, which are distributed in time, indicate that the QPP occurs repeatedly over time. For QPPs detection, both studies (Belloy et al. 2018; Majeed et al. 2011) used the pattern-finding algorithm developed in (Majeed, et al. 2011). Specifically, a window length of 12s (24TR, 1TR=0.5s) was preselected for a mice dataset of 11 male C57BL/6 J mice (22~24 weeks old) (Belloy et al. 2018), which revealed a non-redundant full-size QPP. In comparison, a window length of 20.1s (67TR with 1TR=0.3s) could achieve a non-redundant full-size QPP for a human brain dataset of 6 healthy adults (19~22 yro, 3 females) (Majeed et al. 2011).

## Coactivation Patterns Examples from Literature

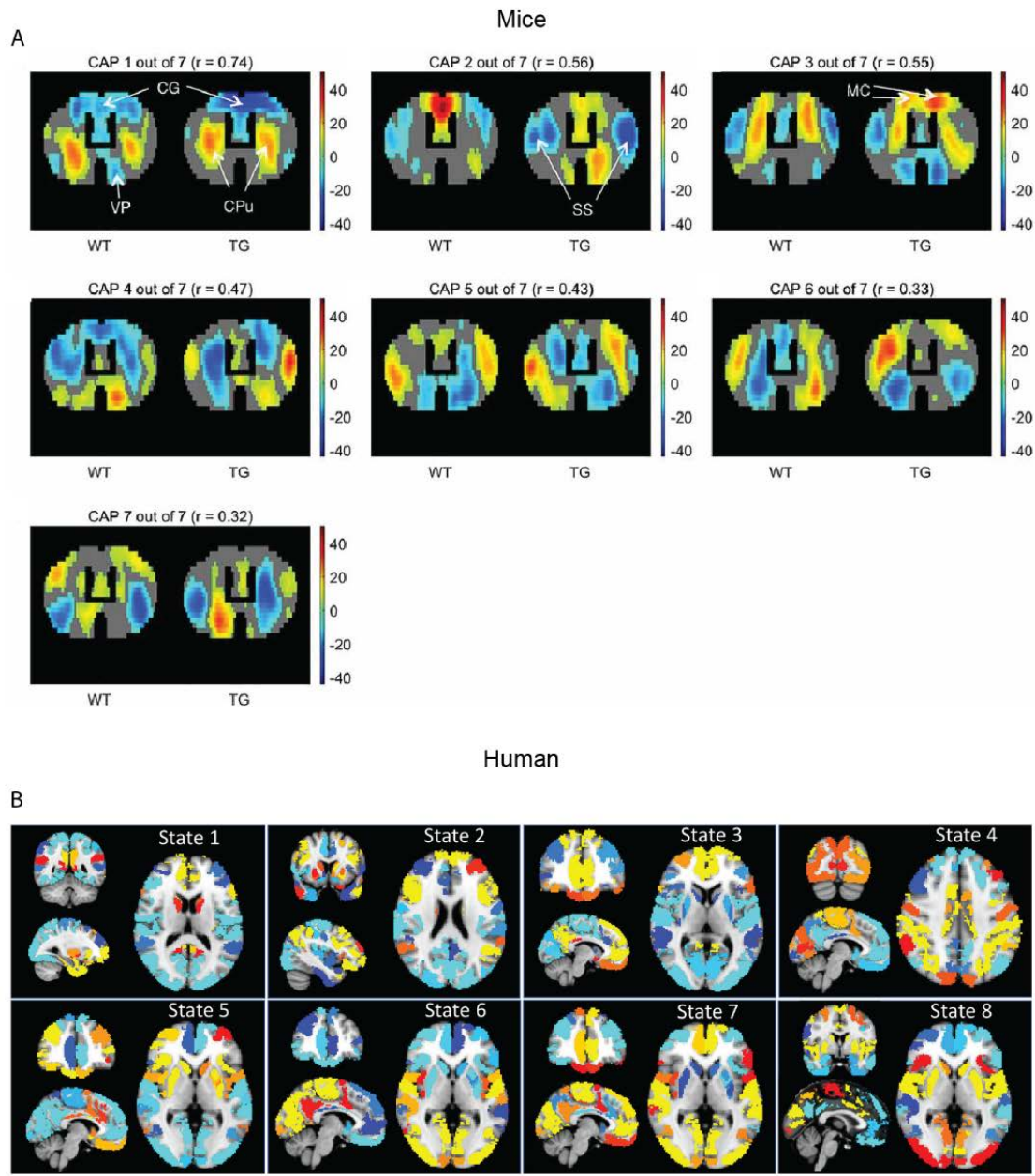

**Figure S5.** Example comparison of coactivations patterns (CAPs) in mice and humans. Brain states determined by CAPs are excerpted from (Adhikari et al. 2021) for mice (A) and from (Janes et al. 2020) for human brains (B). For the mouse study (A), 7 CAP brain states were contrasted between the Alzheimer's disease model of 10 females at 18 months old (referring as the TG group) and 8 age-matched wild-type littermates (referring as the WT group). In particular, significantly (p < 1E-5; Bonferroni corrected) co-activated (T > 0) and co-deactivated (T < 0) brain voxels were tested by one-sample T-test as shown in (A). Distinctions in the following areas were detected between two groups: Cingulate cortex (CG), Caudate Putamen (CPu), Ventral Pallidum (VP), Somatosensory cortex (SS), Motor cortex (MC). For the human study (B), 8 CAP brain states were detected from resting brains of 462 Human Connectome Project subjects (281 females,

22~36 yro) (Van Essen et al. 2013). In particular, to identify the brain states, a K-means clustering analysis of CAPs was performed after an initial PCA dimensionality reduction step. The preselected number of clusters was explored from 4 up to 18, and the clustering solution was evaluated by silhouette scores. The optimal number of cluster K=8 was determined by the highest mean silhouette score. Warm colors represent activation (relative to within-state global average) while cool colors represent deactivation (relative to within-state global average) within each state.